\documentclass[a4paper,12pt]{article}
\usepackage{amsmath,amssymb,epsfig,cite}
\usepackage{multirow}

\newcommand{\preprintnumber}[1]{\vbox{ \baselineskip 14pt \hfill
    \hbox{\normalsize } \\
\hfill \hbox{\normalsize #1} } \vskip 2cm}
\newcommand{\email}[1]{\footnote{email:#1}}

\newcommand{\5}{{\bf 5}}
\renewcommand{\a}{\mathsf{a}}
\renewcommand{\b}{{\sf b}}

\newcommand{\I}{{\rm I}}
\newcommand{\N}{{\cal N}}
\renewcommand{\P}{{\mathbb P}}
\newcommand{\V}{{\cal V}}
\renewcommand{\v}{\check}
\newcommand{\Z}{{\mathbb Z}}
\newcommand{\tr}{{\rm tr\,}}

\begin{document}

\title{ \preprintnumber{KUNS-2282}
$SU(3) \times SU(2) \times U(1)$ Vacua in F-Theory
\\}
\author{
Kang-Sin Choi\email{kschoi@gauge.scphys.kyoto-u.ac.jp}\\
{\it \normalsize Department of Physics, Kyoto University, Kyoto
606-8502, Japan} 
}
\date{}
\maketitle

\begin{abstract}
The Standard Model group and matter spectrum is obtained in vacua of F-theory, without resorting to an intermediate unification group. The
group $SU(3) \times SU(2) \times U(1)_Y$ is the commutant to $SU(5)_\bot \times U(1)_Y$ structure group of a Higgs bundle in $E_8$ and is geometrically realized as a deformation of $\I_5$ singularity. Lying
along the unification groups of $E_n$,
our vacua naturally inherit their unification structure.
By modding $SU(5)_\bot$ out by $\Z_4$ monodromy group,
we can distinguish Higgses from lepton doublets by matter parity. Turning on universal $G$-flux on this part,
the spectrum contains three generations of quarks and leptons, as well
as vectorlike pairs of electroweak and colored Higgses.
Minimal Yukawa couplings is obtained at the renormalizable level.
\end{abstract}

\newpage

\tableofcontents


\section{Introduction}

A good deal of interest has been attracted by model building from
F-theory \cite{F,MV,BJPS}. It provides a geometric description of gauge theories of all the possible Lie groups, automatically incorporating nonperturbative effects \cite{BHV1,BHV2,DW1,DW2,Hayashi:2008ba,Hayashi:2009ge,DW3,Heckman:2010bq,Heckman:2010pv}.

Most studies are devoted to constructions of Grand Unified Theories (GUTs) based on a simple group, because, by relatively simple setup, they allow us to access intermediate routes to the Standard Model (SM) 
\cite{Heckman:2008qt,Andreas:2009uf,Marsano:2009gv,Blumenhagen:2009yv,BHV2,DW2,Hayashi:2009bt,FGUT,Chen:2010ts,Chen:2010tp}.
As a merit, we have more degrees of freedom that are absent in the conventional field theoretic GUTs. For example, we have nontrivial source of four-form flux which breaks gauge symmetry and/or supersymmetry.
In practice, we can obtain the SM from the $SU(5)$ GUT by turning on the flux along the hypercharge $U(1)_Y$ direction without breaking the $U(1)_Y$ itself if some topological conditions are satisfied \cite{DW2,Blumenhagen:2008aw}.
But for this we should take good care of supersymmetry breaking, anomaly cancellation and decoupling of the chiral $X$-boson. Also because of the unification relation, some unobserved interactions are hard to control. 

In this work, we directly build the Standard Model
group and matter contents, without resorting to an intermediate unification \cite{CK}.  One of the obstacle in this approach has been limited understanding on the geometric realization of the SM group, being not a simple group \cite{KM}. Nevertheless the physical structure of unification shall hint us, since the geometric structure reflects that of algebra \cite{Choi}. 

Studies of GUT have revealed that, although the SM group is a combination of some small groups, it is far from arbitrary; The structure of the matter
contents such as charge assignment, quantization and anomaly freedom,
indicates that the unification structure is quite compelling.
The series of exceptional Lie groups of $E_n$-type serves as a promising
route to GUT, harboring the Standard Model with the gauge group \cite{Ramond:1976aw} 
\begin{equation} \label{SMgroup}
 SU(3) \times SU(2) \times U(1)_Y = E_3 \times U(1)_Y.
\end{equation}
Well-known GUT completions include $E_4 = SU(5), E_5= SO(10)$ and $E_6$.
In addition, we learned that a gauge or Higgs bundle background in the internal space
(roughly a gauge field or an adjoint scalar, developing `non-constant' vacuum expectation values (VEVs))
breaks a unification group as well as  makes four dimensional spectrum chiral, enabling gauge--matter unification. In this sense,
$E_7$ or $E_8$ may have a more perfect structure to harbor all the
observed fields including gauge bosons, matters and Higgses into a {\em
single} adjoint \cite{E8eff}. Besides, it is also interesting observation that the structure of neutrino flavor in F-theoretic $SU(5)$ GUT context, we meet $E_8$ as a final group \cite{Heckman:2009mn}.

\begin{figure}[h]
\begin{center}
\includegraphics[height=1.3cm]{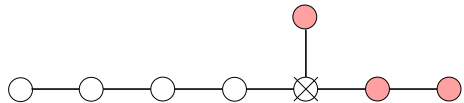}
\end{center}
\caption{The Standard Model group (\ref{SMgroup})
  is the unbroken part of $E_8$, by gauge bundle background of the
  structure group $SU(5) \times U(1)_Y$.}
\label{f:e3}
\end{figure}

F-theory provides natural description of the exceptional groups of $E_n$ series.
Under a mild assumption below, $E_8 \times E_8$ can be a good starting gauge group. 
We shall obtain a low-energy spectrum by breaking it using a background  bundle in the internal space.
In F-theory such background bundle is described by spectral cover \cite{FMW}. 
With the structure groups $SU(9-n)_\bot$, we obtain
$E_n$ group as commutant in $E_8$ \cite{Donagi:2004ia}. 
On top of this, we can also turn on an addition $U(1)$ bundle which
makes the $U(1)$ itself unbroken because an Abelian group commutes to itself
\cite{Blumenhagen:2005ga,Choi}.
Therefore, the background gauge bundle of $SU(5) \times U(1)_Y$
yields the unbroken SM group (\ref{SMgroup}) \cite{CK}. In Figure \ref{f:e3}, this is depicted in $E_8$ Dynkin diagram, by deleting the nodes of structure group.
The GUT relation implies, then, we shall be able to obtain the desired matter fields by the branching of the $E_8$ gaugino. Since the information on the symmetry breaking is contained in the spectral cover, we can
 find the desired matter contents  as well.

Organization: First we briefly discuss how F-theory describes gauge theory with a toy example of $SU(5)$ GUT model.
We can qualitatively understand the gauge group and matter contents purely in terms of algebra, so Section 2 contains such identification. 
A certain property of background Higgs or gauge field is caught by 
the monodromy group, which enables realistic matter contents distinguishing Higgs from lepton doublets. 
Also we can survey Yukawa couplings by gauge invariance.
The model is concretely realized in F-theory in Section 3. First we construct the spectral cover describing the background bundle. This contains all the information, including the parameters used for constructing the SM gauge group, described in terms of singularity of elliptic fiber. Also we identify and calculate homology cycles for the matter curve for the quantitative description.
In Section 4, we turn on $G$-flux to make the theory chiral and calculate the number of generations for the obtained matter fields. We also briefly comment on the requirements on the internal manifold and consequent phenomenologies.
We conclude with an outlook about low-energy phenomenology.

\subsection{Compactification}

Type IIB string theory can be viewed as F-theory compactified on
torus, whose complex structure is identified by
axion-dilaton field $\tau = C_0 + i e^{-\phi}$,
where $C_0$ is Ramond-Ramond scalar
and $\phi$ is dilaton \cite{F}.
The torus is described by the ellptic equation
\begin{equation} \label{ellfiber}
  y^2 + \a_1 xy + \a_3 y = x^3 + \a_2 x^2 + \a_4 x + \a_6
\end{equation}
in space spanned by $x,y \in \P^1$. Perfecting the square and the cube, it becomes Weierstrass form
\begin{equation} \label{origweieq}
 y^2 = x^3 + fx + g.
\end{equation}
It has one complex parameter, related to $\tau$ through Klein's $j$-invariant
\cite{Ap}
\begin{equation} \label{jinv}
{ 4(24f)^3 \over 4f^3+27g^2} = j(\tau).
\end{equation}

To have ${\cal N}=1$ supersymmety in four dimensions, we
compactify F-theory on Calabi--Yau fourfold $X$ \cite{MV,Beetal}.
Because of the above requirement, $X$ is elliptic fibered
over a three-base $B$. The condition of vanishing first Chern class $c_1(X)=0$ requires each
$\a_m$ is a holomorphic section of $K_B^{-m}$, where $K_B$ is the
canonical bundle of $B$.
Along the discriminant locus of (\ref{ellfiber})
\begin{equation}
 S: \{ \Delta \propto 4f^3 + 27g^2 = 0\},
\end{equation}
the torus becomes singular. Using an appropriate coordinates of the
normal bundle to $S$ in $B$, we can regard $\a_m$'s and $f,g$ as
polynomials in that coordinate. We fix the convenient normalization of $\Delta$
later, in (\ref{su5discr}).

The singularities are classified by degrees of $f,g,\Delta$
by Kodaira \cite{MV}. Using intersection theory, the connectedness of a smoothened (blown-up) singularities is the same as that of root system in an $A$-$D$-$E$ algebra, thus they share the same names. However,
in general, when we transport around some base point, the  singularity is effectively reduced to a smaller one, modded out by a monodromy group \cite{Aspinwall:1996nk}. To retain the original $A$-$D$-$E$ symmetry, we need trivial monodromy, which is implemented as so-called the splitness condition that the equation (\ref{ellfiber}) has some factorization structure.
Tate's algorithm includes the check on the
splitness condition just by counting degrees of $\a_m$ \cite{Ta,Beetal}. For $SO(4k+4)$ singularity we need a further information of `complete square' structure. Moreover, this information is crucial in dealing with $SU(n)$ type groups, of which degrees of $f$ and $g$ are zero.
We have displayed the results in Table \ref{t:tate}.

F-theory is a remarkable realization in physics that the singularity  determines the gauge group of the same name in the low energy limit \cite{F}.
We can regard $S$ as a four-cycle where sevenbranes wraps, supporting the  Yang--Mills theory of this gauge group on eight dimensional worldvolume \cite{MV,Beetal}. 
In the perturbative limit, the degree of $\Delta$ is
directly translated in the number of coincident D-branes on $S$.
The gauge fields in Cartan subalgebra come from Kaluza--Klein reduction of three-form field $C_{\it 3}$ in F-theory
\begin{equation} \label{Cartans}
 C_{\it 3} = \sum A^i \wedge \omega_i, 
\end{equation}
where $\omega_i$ are harmonic two-forms corresponding to the positions of  sevenbranes. The ones for non-Abelian directions come from M2
brane wrapping on blown-up cycles of the singularities, whose zero size limit make them massless \cite{DW1}.

In this paper, we additionally require an elliptic K3 fibration structure on $X$ \cite{Choi}. Since elliptic K3 is an elliptic fibration over $\P^1$, the elliptic base $B$ is again $\P^1$ fibration over $S$, and the normal space $\P^1$ has a faithful affine coordinate $z$ of the stereographic projection.
With Calabi--Yau condition, the form of equation (\ref{ellfiber}) is highly restricted, and the maximal group is $E_8 \times E_8$ localized on the opposite poles of the sphere \cite{MV}. Since they are separated \`a la heterotic-M picture and there is no field charged under the both $E_8$'s, we focus only on one of them. We may make use of the other $E_8$ as the origin of supersymmetry breaking.
In fact this description is related to heterotic string on the dual Calabi--Yau threefold, with fiberwise duality between F-theory on K3
and heterotic string on torus \cite{Sen:1996vd}.
Using the same kinds of background bundles for the gauge fields in the internal space, we could obtain similar models\footnote{See also Refs. \cite{Donagi:2004ia}.}.
In F-theory side, however we can concretely obtain the singularity describing the SM group and matter curves localizing the matter fields.

\begin{table}[t]
\begin{center} \small
\begin{tabular}{cccccccccc}
\hline
type  &  group  &  $ \a_1$  &  $\a_2$ &  $\a_3$ &  $ \a_4 $ &  $ \a_6$  & $\Delta$ & $f$ & $g$ \\
\hline
$\I_0 $  &  smooth  & $ 0 $  & $ 0 $  & $ 0 $  & $ 0 $  & 0 &
$ 0$  & $0$ & 0\\
 $\I_1 $  &  $U(1)$  & $0 $  & $ 0 $  & $ 1 $  & $ 1 $  &
$ 1 $  & $1$ & 0 & 0 \\
 $\I_2 $  &  $SU(2)$  & $0 $  & $ 0 $  & $ 1 $  & $ 1 $  &
$ 2 $  & $2$ & 0 & 0 \\
$\I_{2k-1}^{\rm ns}$  & unconven.  &  $0$  & $0$  & $k$  & $k$  &
$2k-1$  & $2k-1$ & 0 &0 \\
 $\I_{2k-1}^{\rm s}$  & $SU(2k-1)$  & $0$  & $1$  & $k-1$  & $k$  & $2k-1$  & $2k-1$ & 0 & 0 \\
$\I_{2k}^{\rm ns}$  & $ Sp(k)$  & $0$  & $0$  & $k$  & $k$  & $2k$  & $2k$ & 0& 0\\
$\I_{2k}^{\rm s}$  & $SU(2k)$  & $0$  & $1$  & $k$  & $k$  & $2k$  & $2k$  & 0 & 0 \\
  $\I\I$  &  ---  & $1$  & $1$  & $1$  & $1$  & $1$  & $2$ & 1 & 1
\\
$\rm III$  & $SU(2)$  & $1$
 & $1$  & $1$  & $1$  & $2$  & $3$ & 1 & 1 \\ $\rm IV^{ns} $  & unconven.  & $1$  & $1$  & $1$
 & $2$  & $2$  & $4$  & 1 & 1 \\ $\rm IV^{s}$  & $SU(3)$  & $1$  & $1$  & $1$  & $2$  & $3$  & $4$ & 1 & 1
\\
 $ \I_0^{*\,\rm ns} $  & $G_2$  & $1$  & $1$  & $2$  & $2$ & $3$ & $6$ & 2 & 3\\
$\I_0^{*\,\rm ss}$  & $SO(7)$  & $1$  & $1$  & $2$  & $2$ & $4$ & $6$ & 2 & 3\\
 $\I_0^{*\,\rm s} $  & $SO(8)^*$  & $1$  & $1$  & $2$  & $2$  & $4$  &
$6$ & 2 & 3 \\
$\I_{2k-3}^{*\,\rm ns}$  & $SO(4k+1)$  & $1$  & $1$  & $k$  & $k+1$
 & $2k$  & $2k+3$  &2 & 3\\ $\I_{2k-3}^{*\,\rm s}$  & $SO(4k+2)$  & $1$  & $1$  & $k$  & $k+1$
 & $2k+1$  & $2k+3$ &2 & 3\\ $\I_{2k-2}^{*\,\rm ns}$  & $SO(4k+3)$  & $1$  & $1$  & $k+1$
 & $k+1$  & $2k+1$  & $2k+4$ &2 & 3\\ $\I_{2k-2}^{*\,\rm s}$  & $SO(4k+4)^*$  & $1$  & $1$
 & $k+1$  & $k+1$  & $2k+1$
 & $2k+4$ &2 & 3 \\ $\rm IV^{*\,ns}$  & $F_4 $  & $1$  & $2$  & $2$  & $3$  & $4$
 & $8$  & 3 & 4 \\ $\rm IV^{*\,s} $  & $E_6$  & $1$  & $2$  & $2$  & $3$  & $5$  &  $8$ & 3 & 4\\
$\rm III^{*} $  & $E_7$  & $1$  & $2$  & $3$  & $3$  & $5$  &  $9$ & 3 & 5\\
$\rm II^{*} $
 & $E_8\,$  & $1$  & $2$  & $3$  & $4$  & $5$  &  $10$ & 3 & 5 \\
 non-min  &  ---  & $ 1$  & $2$  & $3$  & $4$  & $6$  & $12$ & 4 & 6\\
\hline
\end{tabular}
\end{center}
\caption{Singularities  identified by the degrees of coefficients of
  elliptic equation (\ref{ellfiber}) and (\ref{origweieq})
  \cite{Beetal}. Here $k \ge 2$, and the starred ones have a further  condition.}
\label{t:tate}
\end{table}

\subsection{Review on $SU(5)$} \label{sec:su5}

First, we briefly review an $SU(5)$ model. By itself it serves as a baby version of GUT model, but later also we will understand the SM singularity as a deformation of $SU(5)$ singularity.

The $SU(5)$ group is described by $A_4$ singularity or the singular fiber I$_5$.
From Table \ref{t:tate}, the most dominant coefficients of (\ref{ellfiber}) give
\begin{equation} \label{weieq}
 \begin{split}
\a_1 &= -b_5  + O(z), \\
\a_2 &= b_4 z + O(z^2), \\
\a_3 &= -b_3 z^2 + O(z^3), \\
\a_4 &= b_2 z^3 + O(z^4), \\
\a_6 &= b_0 z^5 + O(z^6). \\
 \end{split}
\end{equation}
From the condition on $\a_k$, each coefficient $b_k$ on $S$ belongs
to a section $(6-k)c_1 -t \equiv \eta-kc_1$, where $c_1
\equiv c_1 (S)$ is the first Chern classes of the tangent
bundle of $S$, and $-t \equiv c_1(N_{S/B})$ is of the normal bundle to
$S$ in $B$.\footnote{We will not distinguish the line bundle, its
  first Chern class and Poincar\'e dual divisors, if unnecessary.}
There we took $z$ as a coordinate of the normal space, that is, a section of
$-t$, and the surface $S$ is located at $z=0$, exhibited by the discriminant
\begin{equation} \label{su5discr}
\Delta = b_5^4 R_{\5} z^5 + O(z^6), \quad R_{\5} \equiv b_0 b_5^2 -b_2 b_3 b_5+  b_3^2 b_4.
\end{equation}

Equation (\ref{weieq}) is a deformation of $E_8$ singularity
$y^2=x^3+b_0 z^5$ \cite{MV,DW3,Choi}, thus our $SU(5)$ gauge group is the unbroken part under $E_8 \to SU(5) \times SU(5)_\bot$. The matter contents
are obtained by the branching of gaugino
$$
{\bf 248} \to \bf (24,1) + (1,24) + (5,10) + (10,5) +(\overline
5,\overline{10}) + (\overline{10},\overline{5}).
$$

Group theory guides us to identify the matter curves and Yukawa
couplings. Since there is correspondence between the
group and the geometry, we can relate geometric parameters
$t_i,i=1,\dots,5$ with the weights
of $\bf 5$ of the holonomy group $SU(5)_\bot$ \cite{DW3}.
The actual parameters governing the blowing-up are $b_k$ in
(\ref{weieq}). If the holonomy group is $SU$ or $A$-type, we can
relate  as elementary symmetric polynomials of degree $k$, of $t_i$ \cite{DW3}.
The meaning of this is further studied in terms of monodromy in
Section \ref{s:monodromy}.
For example, the unimodular condition for $SU(5)$ is
\begin{equation} \label{b1}
b_1/b_0 \sim  t_1 + t_2 + t_3 + t_4 + t_5 = 0.
\end{equation}
The sum over cycles is understood as the formal sum of divisors.
Shrinking one of the cycle $t_i \to 0$ is not possible, but the combination
\begin{equation} \label{b5}
 \prod_{i=1}^5 t_i \sim b_5/b_0 \to 0
\end{equation}
is possible. Under this the discriminant (\ref{su5discr}) becomes
$O(z^6)$ implying the gauge symmetry enhancement.
Since $t_i$ corresponds to $\bf 5$ of $SU(5)_\bot$, from the
correlation of $\bf (10,5)$, the matter $\bf 10$ of $SU(5)$ GUT becomes light.
It was the off-diagonal components of the adjoint of locally enhanced
group $SO(10)$, under the branching.
Likewise, the matter $\bf 5$ emerges along $t_i+t_j \to 0$, implying
\begin{equation}  \label{R5}
\prod_{i<j}^5 (t_i+t_j) \sim R_{\bf 5}/b_0^3.
\end{equation}
This is also
understood as local gauge symmetry enhancement
to $SU(6)$.

Each of conditions (\ref{b5}) and (\ref{R5})
specifies a codimension one subspace, curve, on $S$.
We identify the matter spectrum localized on the
curves \cite{Beetal,DW3,Marsano:2009gv}
\begin{equation} \label{mcurves}
\Sigma_{\bf 10} = \{b_5  = 0\} \cap S, \quad \Sigma_{\bf 5} = \{
R_{\bf 5} = 0 \} \cap S.
\end{equation}
From the homology of
$b_i$, they transform as
$$ [\Sigma_{\bf 10}] = \eta- 5c_1, \quad [\Sigma_{\bf 5}] =
3\eta-10c_1. $$

There is further local gauge symmetry enhancement at an
intersection of matter curves. As we shall see shortly, this implies
the existence of the Yukawa coupling.
For the moment, assume that the Higgses have the same quantum number
as the matter fields. The relation (\ref{b1}) indicates that there are couplings
\begin{equation} \label{su5yukawa}
 \begin{split}
{\bf 10 \cdot  \overline 5 \cdot  \overline 5}  &: b_4=b_5= 0 \quad (
t_i )+( t_j +
t_k )+( t_l + t_m  ) = 0, \\
{\bf 10 \cdot 10 \cdot  5} &: b_3=b_5=0 \quad ( t_i )+( t_j)+(
-t_i-t_j) = 0,
  \end{split}
\end{equation}
where all the indices are different. These are respectively $SO(12)$ and $E_6$ symmetry enhancement directions.
Since we have repeated indices, the matter curve for $\bf
10$ should overlap with that for $\bf 5$. In Section \ref{s:mcurve41}
we will see how they are related.
The matter curve is on the six dimensional worldvolume, and its
chiral structure does not distinguish $\bf 5$ and $\bf \overline 5$
yet. This is reflected in the fact that the curve $t_i=0$ is identical
to $-t_i=0$.

The parameters defining matter curves in (\ref{mcurves}) are
precisely the coefficients of the discriminant (\ref{su5discr}), since
vanishing the coefficients implies gauge
symmetry enhancement. Thus they provide two {\em independent} ways of
checking matter curves.


\section{The Standard Model from $E_8$}

The Standard Model matter contents are obtained, relying only on group theoretic analysis, with a caveat of monodromy condition related to a background bundle. It can provide a picture on model building in effective field theory. The realization in F-theory is done in the next section.

\subsection{Matter contents} \label{s:mcontents}

In the perturbative description, branes intersecting at angle localize the matter spectrum at the intersection. It is understood as deformation of a larger stack of branes \cite{Choi:2006hm}; the localized bi-fundamental matter originates from the branching of the adjoint of that larger gauge group. Similarly, if we break a gauge group by deforming the singularity supported on a surface $S$, it gives rise to branching of the gaugino on $S$ and its off-diagonal components become matter fields \cite{KV}. That is,
the decomposition under $G \to H \times B$
\begin{equation} \label{KatzVafa}
 {\rm adj}_G \to ({\rm adj}_H,{\rm 1}) + ({\rm 1, adj}_B) + \bigoplus
 ({\rm R}_H,{\rm R}_B) + {\rm h.c.}
\end{equation}
gives rises to matter field $R_H$, which can be chiral, by the property of Dirac operator under the deforming background.
Here, for $G$ being exceptional group, we can have more than off-diagonal components charged under $H \times B$.
Oppositely, the deformed geometry has local symmetry enhancement direction along which such matter fields are localized. 

In the previous section, we have seen the gauge group always comes from a deformation of $E_8$. For the Standard Model,
the gauge symmetry is from
$$E_8 \to SU(3) \times SU(2) \times U(1)_Y \times SU(5)_\bot $$
so we obtain the matter from the decomposition of its adjoint
\begin{equation*}  \begin{split}
 {\bf 248} & \to
 {\bf (8,1,1)+(1,3,1)+(1,1,1)+(1,1,24)}\\
 & + [X{\bf (3,2,1)}_{-5/6}   + q_\circ {\bf (3,2,5)}_{1/6}+d^c_\circ {\bf (\overline 3,1,10)}_{1/3}   \\
 &  +u^c_\circ {\bf (\overline 3,1,5)}_{-2/3} +
 l_\circ {\bf (1,2,10)}_{-1/2} +e^c_\circ {\bf (1,1,\overline{5})}_{-1}+  \text{c.c.}].
  \end{split}
\end{equation*}
The subscripts are put because this is not fully realistic spectrum until the next subsection.

As in the previous section, we relate five weights of $\bf 5$ of
$SU(5)_\bot$ with the parameters $t_1,t_2,t_3,t_4,t_5$.
Also we denote the weight for $U(1)_Y$ as $t_6$.
The hypercharge $U(1)_Y$ is generated by
\begin{equation} \label{hypergen}
 t_Y = {\rm diag}\textstyle(\frac16,\frac16,\frac16,\frac16,\frac16,-\frac56)
\end{equation}
in the $\{t_1,t_2,t_3,t_4,t_5,t_6\}$ direction. Antisymmetric tensor structure is inherited as sum of the parameters. For example, ${\bf 10}$ of $SU(5)_\bot$ are further distinguished by
$d^c_\circ (\cdot,{\bf 10})_{1/3} \sim t_i+t_j$ and $l^c_\circ(\cdot,{\bf 10})_{-1/2}
\sim t_i + t_j + t_6$.
Accordingly, we can identify the matter fields and the corresponding
localization, as displayed in Table \ref{t:mcurve}.

\begin{table}[t]
\begin{center} \small
\begin{tabular}{ccc} \hline
matter & matter curve & homology on $S$ \\ \hline
$X$  & $t_6 \to 0$  &   $-c_1+y$ \\
$q_\circ$  & $\prod t_i\to 0$   &   $\eta-5c_1$ \\
$d^c_\circ $ & $\prod_{i<j} (t_i+t_j) \to 0$ &   $3 \eta - 10c_1+2y$ \\
$u^c_\circ $ & $\prod (t_i+t_6) \to 0$ & $\eta-5c_1+3y$ \\
$l_\circ$  & $\prod_{i<j} (t_i+t_j+t_6) \to 0$ & $3\eta -10c_1+5y$\\
$e^c_\circ $ & $\prod (t_i-t_6) \to 0$ & $\eta-5c_1+4y$ \\
$\nu^c_\circ$ & $\prod_{i\ne j} (t_i-t_j) \to 0$ & on $C$ \\
\hline
\end{tabular}
\caption{Matter contents and the corresponding curves on $S$. All the
  indices run from 1 to 5 and are different.}
\label{t:mcurve}
\end{center} \end{table}

For the moment we do not distinguish the Higgses $h_{d \circ}, h_{u \circ}^c$ from the
lepton doublets $l_\circ$, since there is only one kind of field for this quantum
number. Then the following relations gives nonvanishing Yukawa couplings \begin{align}
l_\circ h_{d \circ} e^c_\circ = l_\circ l_\circ e^c_\circ  &:
(t_i+t_j+t_6)+(t_k+t_l+t_6)+(t_m-t_6) = 0, \label{lhe} \\
  q_\circ h_{u \circ} u^c_\circ = q_\circ l_\circ^c u^c_\circ  &: (t_i)+(-t_i-t_j-t_6)+(t_j+t_6) = 0,\label{qhu} \\
q_\circ h_{d \circ} d^c_\circ = q_\circ l_\circ d^c_\circ  &: (t_m)+(t_k+t_l+t_6)+(t_i+t_j) = 0,\label{qhd}
\end{align}
where all the indices are different and run from 1 to 5.

A natural observation along $E_8$ embedding is that the SM has another
kind of unification.
The form of the generator (\ref{hypergen}) suggests
that the structure group is unified to $SU(5)_\bot \times
U(1)_Y\subset SU(6)_\bot$. So it is convenient to think of an
intermediate step\footnote{This is one of well-known breaking direction that Dynkin
  diagram cannot describe the subalgebra with correct
  weight vectors \cite{Choi:2003pqa}.} $E_8 \to SU(3) \times SU(2) \times SU(6)_\bot$
\begin{equation*} \begin{split}
 {\bf 248} \to&~ \bf (8,1,1)+(1,3,1)+(1,1,35) \\
 & \bf + (3,2,6) + (\overline 3, 2, \overline 6)
+ (\overline 3,1,15) + (3,1,\overline {15})  +(1,2,20),
  \end{split}
\end{equation*}
consequently
\begin{equation} \label{smspec} \begin{split}
\bf (1,1,35)
 &\to\nu^c_\circ {\bf (1,1,24)} + {\bf (1,1,1)}
  + e^c_\circ {\bf (1,1,5)}_1 + e_\circ {\bf (1,1,\overline 5)}_{-1} \\
\bf (3,2,6) &\to q_\circ {\bf (3,2,5)}_{1/6} + X_\circ {\bf (3,2,1)}_{-5/6} \\
\bf (\overline 3,1,15) &\to d^c_\circ {\bf (\overline 3,1,10)}_{1/3}+ u^c_\circ {\bf (\overline 3,1,5)}_{-2/3} \\
\bf (1,2,20) &\to l_\circ {\bf (1,2,10)}_{-1/2} + l^c_\circ {\bf (1,2,\overline
  {10})}_{1/2}.
  \end{split}
\end{equation}
In the last line, the counting of $\bf 10$ and $\bf \overline{10}$ agrees
thanks to the tracelessness of $SU(6)_\bot$
\begin{equation} \label{lepindex}
 t_i+t_j+t_6 =-t_k-t_l-t_m.
\end{equation}
For example ${\bf 15}$ is expressed as $t_i +t_j + t_j$ with all the indices different.

The remaining combinations are moduli from the $SU(5)_\bot$
gaugino. Although this gauge symmetry is broken, in four dimensions
hypermultiplet part can survive. Because they are SM singlet, all
candidates of right-handed neutrinos
$\nu^c_\circ: {\bf 24}_0: t_i - t_j \to 0.$

\subsection{Modding out by monodromy} \label{sec:z4monodromy}
\label{s:monodromy}

In (\ref{KatzVafa}), the subgroup $B$ is broken by the background bundle , so the visible quantum numbers are those of $H$. The problem is how to count the quantum number for $H$.
Recall that we could easily understand the property of the parameters $b_k$ in $SU(5)$ singularity (\ref{weieq}), by relating with elementary symmetric polynomials of order $k$, made of $t_i$ \cite{DW3}
\begin{equation}
b_k /b_0 \sim s_k , \quad \prod_{i=1}^5(x+t_i) = \sum_{k=0}^5 s_k x^{5-k}. \end{equation}
We related $t_1 ,\dots t_5$ with weights of $\bf 5$ of the structure group $SU(5)_\bot$ and $t_6$ with generator of $U(1)_Y$
\begin{equation}
 a_1/a_0 \sim t_6.
\end{equation}

This identification implicitly involves the notion of the monodromy of the structure group. Since the `physical parameters' are $b_k$ (and $a_0,a_1$), we do not distinguish different $t_i$. In effect, for every combinations of $b_k$, we mod out by $S_5$ permutating all the elements $t_1, \dots, t_5$. The connected ones by $S_5$ form an orbit and are treated as the identical curve \cite{Hayashi:2009ge}. As an example, consider $l_\circ {\bf (1,2,10)}_{-1/2}$ transforming as $\bf 10$ of $SU(5)_\bot$.  After modding out, these ten elements form the closed orbit
\begin{equation}
 \{ t_i + t_j + t_6 \ |  \text{ $i,j$ run over 1,2,3,4,5 and different} \}
\end{equation}
and are counted as just one field.

If we have $\Z_5$ monodromy, we have two kinds of
such doublet,
\begin{equation*}
\begin{split}
[123] = \{t_1+t_2+t_3,t_2+t_3+t_4,t_3+t_4+t_5,t_1+t_4+t_5,t_1+t_2+t_5\}, \\
[134] = \{t_1+t_3+t_4,t_2+t_4+t_5,t_1+t_3+t_5,t_1+t_2+t_4,t_2+t_3+t_5\}.
\end{split}
\end{equation*}
Here we use
the notation including only one element in the orbit. It follows, for instance, $[123]=[125]$ and $[134]=[124]$.
For the negative sign in front of $t$, we put bar on the index. For example by Eq.~(\ref{lepindex}), $[126]=[\bar 3 \bar 4 \bar 5]$.
All of these are insufficient to distinguish three kinds of fields
with the same quantum number, namely $l,h_d$ and the conjugate of
$h_u$, then forming $\mu$-term or Yukawa coupling mediating lepton number
violating process.
This leads us to consider a smaller group.

The next candidates are of degree four, that is, transitive subgroups of $S_4$. 
First, consider $\Z_4$, which is generated by the cyclic permutation of the four elements, say $t_1,t_2,t_3,t_4$, singling out $t_5$, without loss of generality.
We obtain the candidates for matter curves
\begin{equation} \begin{split}
q & : [1],[5] \\
X       &: [6] \\
d &: [12], [13], [15] \\
u^{c} &: [16], [56] \\
l,h_d,h_u   &: [\bar 1\bar 2\bar 3], [\bar 1\bar 2\bar 5], [\bar
  1\bar 3\bar5]  \\
e^{c} &: [1\bar6], [5\bar6].
  \end{split}
\end{equation}
\begin{table}[t]
\begin{center}
\begin{tabular}{cccc} \hline
matter & matter curve & homology on $S$ & $M$ \\ \hline
$X$  & $t_6 \to 0$  & $-c_1$ & $0$ \\
$q$ & $\prod t_i\to 0$    & $\eta-4c_1-x$ & $1$ \\
$q'$ & $t_5\to 0$    & $-c_1+x$  & $-4$ \\
$d^c $ & $\prod(t_i+t_5) \to 0$ & $ \eta - 4c_1+2x$ & $-3$ \\
$D_1^c$ & $(t_1+t_3)(t_2+t_4) \to 0$ & $\eta - 2c_1-x$ & $2$ \\
$D_2 $ & $\prod(t_i+t_{i+1}) \to 0$ & $\eta - 4c_1-x$ & $2$ \\
$u^c $ & $\prod(t_i+t_6) \to 0$ & $\eta-4c_1-x$ & $1$ \\
$u' $ & $t_5+t_6 \to 0$ & $-c_1+x$ & $-4$ \\
$h_u^c    $  & $\prod(t_i+t_{i+1}+t_6) \to 0$
 & $\eta -4c_1-x$ & $2$ \\
$h_d$  & $(t_1+t_3+t_6)(t_2+t_4+t_6) \to 0$
 & $\eta -2c_1-x$ & $2$ \\
$l    $  & $\prod(t_i+t_5+t_6) \to 0$
 & $\eta -4c_1+2x$ & $-3$ \\
$e^c $ & $\prod(t_i-t_6) \to 0$  & $\eta-4c_1-x$ & $1$ \\
$e' $ & $t_5-t_6 \to 0$  & $-c_1+x$ & $-4$ \\
$\nu^c$ & $\prod(t_i-t_5) \to 0$ & $\eta-4c_1+2x$ &  $5$ \\
$\nu_1$ & $(t_1-t_3)(t_2-t_4) \to 0$ & & 0 \\
$\nu_2$ & $\prod(t_i-t_{i+1}) \to 0$ & & 0 \\
\hline
\end{tabular}
\caption{Matter contents, modded out by $\Z_4$ monodromy.
All the indices take different value in $\Z_4=\{1,2,3,4\}$.
The primed ones are charged exotics, to be decoupled. Note that they
have always odd $M+2s$ charges.}
\label{t:mcurve41}
\end{center} \end{table}
To have lepton Yukawa coupling (\ref{lhe}), $h_d$ and $l$ must not share
$t_5$. In this paper, we choose
\begin{equation} \label{second41}
 h_u: [125], \quad h_d: [\bar 1\bar 3\bar 5], \quad  l: [\bar 1\bar
   2\bar 3],
\end{equation}
resulting in
\begin{equation}
\quad e^c: [1 \bar 6], \quad u^c:[16], \quad q:[1],\quad d^c:[15].
\end{equation}
We can see that, although not so trivial, two sets under the exchange $[12]
\leftrightarrow [13]$ {\em and} $[125] \leftrightarrow [135]$ also,
are equivalent. However for the remaining color triplets, we still
have two choices for $[12]$ or $[13]$.
We proceed with the choice
\begin{equation}
 q':[5],\quad u':[56], \quad e':[5 \bar 6], \quad D_1^c :[13], \quad D_2:[12].
\end{equation}
They are unobserved charged exotic particles, thus should be
decoupled. We summarized these in Table \ref{t:mcurve41}. In the context of  $SU(5)$ unification, we will call $D_1$ and $D_2$ colored Higgses.

The other choice is
\begin{equation} \label{first41}
h_d: [\bar 1\bar 2\bar 3], \quad l: [\bar 1 \bar 2 \bar 5], \quad h_u: [135]
\end{equation}
and the $\mu$-term is forbidden. Following (\ref{lhe})-(\ref{qhd}) we can
determine all the rest as
\begin{equation*}
e^c:[1 \bar 6],\quad u^c: [16], \quad q: [1], \quad d^c: [12].
\end{equation*}

The monodromy group should be smaller than $S_4$
to distinguish lepton doublet and Higgs
doublets, for which at least we need three distinct orbits.
One can check that modding out by $D_4$ monodromy has the same
effect as $\Z_4$ \cite{Marsano:2009gv}. The only other proper subgroup of $S_4$ is
$V=\{1,(12)(34),(13)(24),(14)(23) \}$
group, where the parenthesis denote the permutations of $t_i$'s with the indicated indices. The only difference is that each lepton doublet and down type
quarks has one more kind of representation
\begin{equation}
 d: [12], [13], [14], [15], \quad
 l: [\bar 1\bar 2 \bar 3], [\bar 1 \bar 2 \bar 5], [\bar 1 \bar 3 \bar
   5], [\bar 1 \bar 4 \bar 5].
\end{equation}
However, we also see that the extra fields, say $[14]$ and $[\bar 1 \bar 4 \bar 5]$, having homology $-2c_1$ on $S$, will be
projected out for the same reason for $X,q',u',e'$.
Then, as the notation suggests, one can check that the nonvanishing
Yukawa couplings are identical to previous cases of $\Z_4$ and $D_4$. Thus any degree-four transitive discrete group except $S_4$ gives the same model.

We should account for the observed number of three generations of fermions. The multiplicity of fermions is accounted by the number of zero modes of
Dirac operator in the internal dimension.
We expect that each of observed fermions shall have three zero modes, which is the approach we take in what follows. Alternatively we can seek a matter assignment distinguishing different generations \cite{DP}. For example, instead of naming $l,h_u^c,h_d$ we may say they are three different generations of $l$, giving rise to flavor dependent quantum numbers. This is not possible for the quark sector in this simplest setup. Modding out by smaller group may realize this but at the price of opening up extra $U(1)$ gauge symmetries.

\subsection{Yukawa coupling} \label{s:Yukawa}

The low-energy effective worldvolume theory can be obtained as in conventional dimensional reduction of super Yang--Mills theory.
Effects from the elliptic fiber are amended by topological twist,
redefining the holomomies of the elliptic fiber and $U(1)$ part of
K\"ahler manifold $S$ \cite{BHV1,Conlon:2009qq}.
Since the matter fields are obtained by branching of gaugino $\lambda$,
its twisted action, corresponding to the covariant derivative $\overline \lambda [A,\lambda]$ in supergravity, gives rise
to Yukawa coupling \cite{BHV1}. Its existence is
guided by gauge invariance, whose condition is vanishing the sum of weight
vectors.
Since the matter identity is specified by the matter curve, we check
it by the condition on structure group.

\begin{table}[t] \begin{center} \begin{tabular}{|c|c|c|}
\hline coupling & condition & note \\ \hline \hline
$l h_d e^c$ & $[156] [246] [3 \bar6]$ & \\
$  q h_u u^c$ & $ [1] [\bar 1 \bar 2 \bar 6] [26] $ & MSSM \\
$q h_d d^c$ & $ [1] [246] [35] $ & superpotential\\
$l h_u \nu^c$ & $[126][\bar 1 \bar 3 \bar 6][3 \bar 2] \label{lhnu}$ &\\
\hline
 $\nu_1 \nu^c_1$ & $[1 \bar 2] [\bar 1 2]$ & $[1\bar 2] \ne
    [\bar 1 2]$ \\
     $\nu_2 \nu^c_2$ & $[1 \bar 2] [\bar 1 2]\label{Maj}$ & $[1\bar 3] \ne
    [\bar 1 3]$ \\
     \hline
$X h_u d^c$ & $[6][\bar 1 \bar 2\bar 6][12] \label{Xhd}$& nonvanishing
    \\
 \label{qphd}
$ q^{\prime} h_d^c D_1^c$ & $[5][\bar 1\bar 3\bar 5][13]$& but $X$
 and $q'$  \\
$ q' h_u D_2 $& $[5] [\bar 1 \bar 2 \bar 5][12]$&  are absent  \\ \hline
    \end{tabular}
  \end{center}
\caption{Nonvanishing Yukawa couplings, using representative notation.} \label{t:Yukawa}
\end{table}

The nonvanishing Yukawa couplings are displayed in Table \ref{t:Yukawa}.
Here we also use the representative notation. For example, $[1][246][35]$ for $qh_d d^c$ in (\ref{qhd}) means the triple intersection
$$ t_m = t_k+t_l+t_6 = t_i+t_j = 0, $$
with the redundant condition
$$ (t_m)+(t_k+t_l+t_6)+(t_i+t_j) = 0. $$
In each line, we have more than one of writting {\em the
same} coupling due to the relation (\ref{lepindex}).
For example we can rewrite the same relation as $[1][\bar 1 \bar 3 \bar 5][35]$
\begin{equation}
(t_m)+(-t_m-t_i-t_j)+(t_i+t_j) = 0, \ \epsilon_{ijm} \ne 0.
\end{equation}
These will be related to distinct GUT extension.

The monodromy choice distinguishes the SM singlets
\begin{equation}
 \nu^c : [1\bar 5], \quad \nu_1: [1 \bar 2], \quad \nu_2: [1
   \bar 3].
\end{equation}
We see $\nu^c$ becomes right-handed Dirac neutrino forming mass terms $l h_u \nu^c$, as in Table \ref{t:Yukawa}. On the other hand, since $t_i-t_j$ is distinguished from
$t_j-t_i$, $\nu_M,M=1,2$ can have self-coupling $\nu_M \nu^c_M$,
thus a heavy Majorana mass can generate see-saw mechanism.

Depending on choice of monodromy and particle identity there can be the couplings in the last row in Table \ref{t:Yukawa}.
There are couplings involving $X$ or $q'$, but obviously there is no problem if they do not exist, or both be superheavy. It turns
out that, in the later configuration, we can have {\em none} of them in {\em four dimensional} effective theory.

It follows that there is no lepton or baryon number
violating terms
\begin{equation}
h_u h_d, \ lh_u,\ lle^c,\ lqd^c,\ u^c d^c d^c,
\end{equation}
and so on, at the renormalizable coupling.
In the first choice (\ref{first41}), the lepton doublet has odd $U(1)_M$ charge
whereas in the second case (\ref{second41}), only Higges has even $U(1)_M$ charge.

There can be baryon and/or lepton number violating operators in higher order couplings. The configuration allows dimension five operator, for instance,
\begin{equation}
 \frac{c}{M_F}qqql
\end{equation}
present in supersymmetric standard model. If we take F-theory scale
$M_F$ around the Planck scale, the parameter $c$ should be smaller than the unobserved bound $10^{-5}$. At present we have no way of calculating tree-level non-renormalizable couplings, since we derive the Yukawa couplings from the dimensional reduction of the effective theory. Usually it is known that higher order couplings are sufficiently suppressed by instanton-like corrections as a form of area-law in stringy worldsheet \cite{Abel:2003yx}, Euclidian brane \cite{Font:2009gq}, and effective field theory \cite{Abe:2009dr} calculations exhibit similar behavior.
We do not have other dimension-five lepton number violating operators $llhh$ or $u^c u^c u^c e^c$.

\subsection{Matter parity and unification relations}

The fact that many couplings are forbidden can be tracked
to $U(1)_M$ charge generated by
\begin{equation} \label{u1bminusl}
 t_M =  {\rm diag}(1,1,1,1,-4,0)
\end{equation}
in the basis $\{t_1,t_2,t_3,t_4,t_5,t_6\}$. This is a continuous version
of matter parity, to be combined with $R$-symmetry. This is nothing but the commutant to $SU(5)$ in $SO(10)$, with which we have the
standard relation with the baryon minus the lepton number $(B-L)$
\begin{equation} \label{bminusl}
 M =  4t_{3R} + 3(B-L),
\end{equation}
where we assign $t_{3R}$ eigenvalues $\{\frac12,-\frac12\}$
respectively for the up and down type counterparts of right-handed fermions
\begin{equation}
 t_{3R} = \textstyle {\rm diag}(0,0,0,0,-\frac12,\frac12).
\end{equation}
The matter curves show this symmetry is manifest under the exchange of $t_5$ and $t_6$.
For example, the exchange symmetry for $h_u$ and $h_d$ is the exchange symmetry between $[\bar 3 \bar 4 \bar 6]=[125]$ and $[136]$. We can check similarly for the $D_1$ and $D_2$ exchange.
The famous left-right relation follows
\begin{equation}
 \textstyle Y = \frac16( M -10t_{3R}) = \frac12(B-L)-t_{3R} .
\end{equation}
If a combination of $U(1)_M$ and $U(1)_R$ is diagonally broken to $\Z_2$
symmetry, for example by vacuum expectation value of even-charged
scalar, this becomes nothing but {\em $R$-parity.}

We had singlets naturally recognized as right-handed neutrinos.
Being charged under $U(1)_M$, $\nu^c$ should be Dirac
neutrino. It is the $SU(5)$ singlet inside $\bf 16$ of $SO(10)$
GUT. On the other hand, $\nu_1$ and $\nu_2$ are neutral under
all the symmetries, so it can have self-coupling to have Majorana
mass. They are singlets outside $\bf 16$ but later to be interpreted
inside $\bf 27$ of $E_6$.

We can easily see the gauge symmetry enhancement direction means the unification. For example
$SU(5)$ GUT relation in (\ref{su5yukawa}) is restored in the limit
$t_6 \to 0$. As in Table \ref{t:mcurve},
the distinct curves for $(q,u^c,e^c)$ reduce to the same one $\prod t_i \to 0$, implying that
they are unified to a single multiplet $\bf 10$. The ones for $(l,d^c)$ reduces to $\prod (t_i+t_j) \to 0$.
Accordingly, (\ref{lhe}), (\ref{qhd}) is unified to $\bf 10
\cdot \overline 5 \cdot \overline 5 $ in
(\ref{su5yukawa}) and (\ref{qhu}) goes to $\bf 10 \cdot 10 \cdot
5$. Right-handed neutrino is neutral under this $SU(5)$.

The famous unification limits are shown in Table \ref{t:gutenh}.
\begin{table}[h]
\begin{center} \begin{tabular}{cccc} \hline
 unification & group & limit & unhinggsing \\ \hline
 Georgi--Glashow  & $SU(5)$ & $t_6 \to 0$ & $X$ \\
 flipped $SU(5)$ & $SU(5) \times U(1)$ & $t_5 \to 0$ & $q'$ \\
 $SO(10)$ & $SO(10)$ & $t_5 \to 0, t_6 \to 0$ & $X,q',u',e'$\\
 left-right & $SU(3) \times SU(2)_L \times SU(2)_R$ & $t_5-t_6 \to 0$ & $e'$ \\ \hline
\end{tabular}
\caption{Various enhanced GUTs (not local enhancement).}
\end{center} \label{t:gutenh}
\end{table}
Later we shall see that realization of these as local unification group will hint to decouple exotic matters. The Pati--Salam enhancement to $SU(4) \times SU(2)_L \times SU(2)_R$ is not possible. The limit $t_5+t_6 \to 0$ unifies $q$ and $l$ into $\bf (4,2,1)$, but there is no $\bf (4,1,2)$ unification. This is because Pati--Salam model cannot be obtained by a $(S)U(n)$ background bundle.


\section{Realization in F-theory}

In the preceding sections, we identified the matter contents purely in terms
of quantum numbers of the structure group, modded out by monodromies.
In the following we will see how they are realized in F-theory.
Expressing them in terms of group theoretical weights reveals certain
relations among matter curves \cite{DW3,Marsano:2009gv}. In case
of $SU,SO$ or $Sp$ structure group, it is most
extensively described by spectral covers \cite{FMW,Do97,Donagi:2004ia}. So we first construct the spectral cover, from which we also learn all the information on the parameters used in the singularity describing the SM group.

\subsection{The spectral cover for $SU(5)_\bot \times U(1)_Y$}

The  commutant to the SM group in $E_8$ is $SU(5)_\bot \times
U(1)$. In the heterotic dual language, it is the structure group of the
background gauge bundle thus broken.
Coming back to the effective field theory limit on the eight
dimensional worldvolume, this is translated to a background Higgs bundle. It is roughly non-constant VEVs of the adjoint Higgs $\varphi$,
with the eigenvalues as sections of the canonical bundle $K_S$
\cite{BHV1,DW3}. It is a scalar part of vector multiplet with topological
twist. In the perturbative
case, taking $T$-duality in the normal direction to $S$,
its eigenvalues become the positions of D-branes relative to $S$.
In our theory, we neither have perturbative D-branes nor the spacetime is flat, there is no such duality. However it is convenient to imagine a dual of $\varphi$ in the sense that eigenvalues in the group space becomes the positions of some seven-branes in a certain space. Such generalization is described as follows by spectral covers \cite{FMW}.
We can think of a polynomial whose roots are the eigenvalues $t_i$.
For $SU(5)_\bot \times U(1)$ we have respectively 5 and 1 eigenvalues, so we
consider an equation having them as the roots
\begin{equation} \label{su5xu1eq}
0 = (a_0s + a_1)(b_0 s^5 + b_1 s^4 + b_2 s^3 + b_3 s^2+ b_4 s + b_5)  \equiv F_X \cdot F_5.
\end{equation}
Our surface $S$ is defined to be located at $\{s=0\}$.
Once we identified the SM group along the $E$-series, the parameters in second
factor, the $SU(5)_\bot$ part in Eq. (\ref{su5xu1eq}), should be that of $SU(5)$ GUT. Then, still  $b_i$ are sections of $\eta-kc_1$ on $S$, and
consequently, the coordinate $s$ should be a section of $-c_1=K_S$.
It intersects the first factor at the matter curve $a_1 = 0$, and the second factor at $b_5 = 0$.
For this reason the spectral surface is sometimes called the flavor brane.

To recycle the parameters $b_k$ in equations (\ref{weieq}), we make the ambient space $\v Z$ as some fibration over $S$.
Later we need an information at the infinity
$s\to \infty$, we make $\v Z$ compact by projectivization
\begin{equation} \label{threefold}
 \v Z= \P({\cal O} \oplus K_S) \to S,
\end{equation}
where ${\cal O}$ is a trivial bundle on $S$. It is generated by two
divisors $\sigma$ (zero section) and $\sigma_\infty \equiv \sigma + c_1$, which are disjoint
\begin{equation} \label{disjoint}
 \sigma \cap \sigma_\infty =0 \quad \Longleftrightarrow \quad \sigma \cap \sigma =
 -\sigma \cap c_1.
\end{equation}
The intersection $\cap$ is done between divisors in $\v Z$.
We introduce global coordinates $U$ and $V$ in the project space, as section of $\sigma$ and $\sigma_\infty$. They respectively correspond to above $s$ and $1/s$. Now $S = \{U=0\}$ is the vanishing locus of the section $\sigma$, and
$\sigma_\infty$ has coordinate $V$ such that $V=0$ corresponds
to $s \to \infty$.
Using the splitting principle, the first Chern class is
\begin{equation} \label{c1X}
 c_1(\v Z)= c_1 + \sigma + \sigma_\infty = 2 \sigma+2c_1,
\end{equation}
which does not always vanish. Now the spectral cover is a variety in $\v Z$
\begin{equation}\label{su5xu1cover}
C_X \cup C_5:  (a_0 U+a_1 V)(b_0 U^5+b_1 U^4 V+ \dots + b_5 V^5) = 0, \end{equation}
which is a virtual sixfold covering of $S$ \cite{DW3}.

We might have
a freedom to choose $a_0 \sim y$ for a certain $y \in H_2(S,\Z)$, but it will turn out
to be trivial.
Here we {\em defined} an auxiliary coefficient
\begin{equation} \label{su6traceless}
 b_1 = - \frac{a_1}{a_0} b_0
\end{equation}
which is a section of $\eta - c_1$.
It makes sense since the structure group can be embedded into
$SU(6)_\bot$, whose traceless condition is nothing but (\ref{su6traceless}). For this reason such group is sometimes called $S[U(6)_\bot \times U(1)_Y]$.
We obtain $SU(5)$ GUT by local gauge symmetry
enhancement $t_6 \sim a_1/a_0 \to 0$ on the matter curve of $\v Z$, which should
reduce to the tracelessness of $SU(5)_\bot, b_1 \to 0$.

Letting the projection $\pi: \v Z \to S$, the homology classes of the spectral surfaces in (\ref{su5xu1cover}) are
\begin{equation}
 [C_X] = \sigma+ \pi^* y,\quad [C_5] =5 \sigma+ \pi^* \eta.
\end{equation}
On $S$, their intersections give the matter curves for $X$ and $q$, respectively
\begin{equation}  \label{fundmattercurves}
 [\Sigma_X] = [C_X \cap \sigma]_S = -c_1 +  y,\quad [\Sigma_{q_\circ}] = [C_5 \cap \sigma]_S = \eta -5 c_1  ,
\end{equation}
using the relation (\ref{disjoint}).

\subsection{The $SU(3) \times SU(2) \times U(1)_Y$ singularity}

The Standard Model group $SU(3)\times SU(2)\times U(1)_Y$ is
not a simple group, so not the entire group appears in Table \ref{t:tate}. 
To find it, an important clue is the unification structure: the group and the matter contents are not arbitrary but embeddable to a larger group with simpler structure.
Since the Lie algebra has the same connectedness with the singularity sharing the same name, so we obtain a singularity for a semisimple group by
deforming that of unifying simple group \cite{Choi,CK}.
The minimal choice leads us the $SU(5)$ curve discussed in Section \ref{sec:su5}.

\subsubsection*{Hint from the local spectral cover}

First, we ask what parameters are needed. We see that
its commutant group $SU(5)_\bot \times U(1)_Y$ is described by spectral cover
(\ref{su5xu1cover}),
\begin{equation} \label{SMcover}
 b_0 U^6+ (b_2+a_1 b_1 ) U^4 V^2+(b_3+a_1 b_2) U^3 V^3 +
(b_4+a_1 b_3) U^2 V^4 + (b_5+a_1 b_4) U V^5+ a_1 b_5 V^6 = 0,
\end{equation}
where we set $a_0 = 1$, to be justified later by six dimensional anomaly consideration. Then we can reuse the
parameters $b_k$ for the $SU(5)$ in (\ref{weieq}).
As a result, we introduce
one new parameter $a_1$, a section of $-c_1$.  Unlike $SU(6)_\bot$ spectral cover, we do not have any `elementary' parameter as a section of $-t$ like $z$, but only the combination $a_1b_5$, which does not harm ampleness of the base space $S$ \cite{DW3}.

We expect the SM group is the deformation of $SU(5)$ GUT group to which the symmetry is restored in the $a_1 \to 0$ limit. Since $SU(5)_\bot$ is the special case of $SU(6)_\bot$, we want to preserve the combination in the coefficients in (\ref{SMcover}). 
Now, if only one simple group is describable at once at one supporting surface, we require that the singularity should locally look like either $SU(3)$ or $SU(2)$. Either choice should be physically equivalent, so we take the former. 
Also we assume, there can be a back-reaction of the $SU(3)$ brane against the $SU(2)$ brane, while the deformation maintaining the center-of-mass at $S$, so we choose a different coordinate $z'$ than $z$. Requiring the degrees of $(\a_1,\a_2,\a_3,\a_4,\a_6,\Delta)$ to be respectively
$(0,1,1,2,3,3)$, from Table \ref{t:tate}, we need deformations in $\a_3,\a_4,\a_6$. Therefore we are led to \cite{CK}
\begin{equation}
 \begin{split} \label{smsurface}
\a_1 &= -(b_5 + b_4 a_1) + O(z'),\\
\a_2 &= (b_4 + b_3 a_1)z' + O(z^{\prime 2}),\\
\a_3 &= -(b_3 + b_2 a_1)(a_1 b_5+ z')z' + O(z^{\prime 3}), \\
\a_4 &= (b_2 + b_1 a_1)(a_1 b_5 +  z')z^{\prime 2} + O(z^{\prime 4}), \\
\a_6 &= b_0 (a_1 b_5 +  z' )^2 z^{\prime 3} + O(z^{\prime 6}),
 \end{split}
\end{equation}
which most simply describe the desired singularity at $\{z'=0\}.$ Again recall that we defined $b_1 = -a_1 b_0$ in (\ref{su6traceless}). We have freedom to choose the scaling of $z'$ thus the choice of the coefficient of $a_1 b_5$.

By degree counting, the {\em generic} singularity is $\I_3^{\rm s}$ for $SU(3)$. However the parameters are specially tuned, thus the actual singularity is larger \cite{Choi}.
Obviously they are the deformations of (\ref{weieq}) by adding lower
order terms in $z$, that is, exact up to $O(z^{\prime 5})$. The discriminant takes the form
\begin{equation} \label{SMdiscr}
 \Delta =  (b_5+a_1 b_4)^3 P_X^2 P_{q_\circ}^2 P_{d^c_\circ} P_{u^c_\circ} z^{\prime 3}  + P_{q_\circ} P_X Q_4
 z^{\prime 4} + O (z^{\prime 5}).
\end{equation}
The parameters are shown in Table \ref{t:mequation}, and $Q_4$
parameterizes the rest of the coefficient in $z^{\prime 4}$ term. One can obtain them by
brute force calculation by plugging the explicit form
(\ref{smsurface}) into discriminant and factorizing it.
We can verify that it {\em agrees} with the matter curves obtained by group
theoretical reasoning in Section \ref{s:mcurve41}.

\begin{table}[t]
\renewcommand{\arraystretch}{1.3}
\begin{center} \begin{tabular}{|c|c|c|} \hline
matter & equation in terms of parameters & of weights \\ \hline
$P_X$ & $a_1$ &$t_6$  \\  \hline
$P_{q_\circ}$ & $b_5$ & $\prod t_i$ \\ \hline
$P_{u^c_\circ}$ & $b_2 a_1^3 + b_3 a_0 a_1^2 + b_4 a_0^2 a_1 + b_5 a_0^3$
& $ \prod (t_i+t_6)$ \\
\hline
$P_{d^c_\circ}$ &
 \parbox{8cm}{ $a_0^2b_0b_5^2-a_0^2b_2b_5b_3+
 a_0^2b_4b_3^2 + 2a_0a_1b_5b_0b_4$ \\ $
 -a_0b_2^2b_5a_1+a_0b_2a_1b_4b_3 +b_0a_1^2b_4^2$
 } &
 $\prod(t_i+t_j)$
\\ \hline
$P_{l_\circ}$ &
\parbox{10cm}{
 $ a_1^5
b_5b_0^2-a_1^3b_5a_0^2b_2b_0-2a_1b_5a_0^4b_0b_4-3a_1^2b_5a_0^3b_0b_3$ \\
$+a_0^3b_0a_1^2b_4^2 +a_0^5b_0b_5^2+a_0^4
b_2a_1b_4b_3-a_0^5b_2b_5b_3+a_1^3b_0b_4a_0^2b_3$\\
$+a_0a_1^4b_0 b_4b_2 +a_0^5b_4b_3^2+a_0^4a_1
b_3^3+2a_0^3a_1^2b_3^2b_2+a_0^2a_1^3b_3b_2^2$
} &

$\prod(t_i+t_j+t_6)$
 \\ \hline
$P_{e^c_\circ}$ &  $-2 b_1 a_1^4 +b_2 a_0 a_1^3- b_3 a_0^2 a_1^2 +b_4 a_0^3
 a_1^2 - b_5 a_0^4 $ &
$\prod(t_i-t_6)$
\\ \hline
$(P_{\nu^c_\circ})$ & over the bulk of $C$ &  $\prod_{i\ne j} (t_i-t_j) $ \\
\hline
  \end{tabular}
\caption{Defining equation for the matter curves. See Table
  \ref{t:mcurve}. Here $S_5$ monodromy is used and similar expressions are also found for $\Z_4$ case, as in Table \ref{t:mcurve41}.}
\label{t:mequation}
\renewcommand{\arraystretch}{1}
\end{center}
\end{table}

From the mechanism (\ref{KatzVafa}),
vanishing matter curves are responsible for gauge symmetry enhancements. Since $X$ and $q$ have nonabelian charges $\bf (3,2)$, either $P_X = a_1=0$ or $P_{q_\circ}= b_5 =0$ enhances the singularity to $O(z^{\prime 5})$ which is $\I_5^{\rm s}$ and the gauge symmetry is enhanced to $SU(5)$. Since $u^c_\circ$ and $d^c_\circ$ has charged only under $SU(3)$, for $P_{u^c_\circ}=0$ or $P_{d^c_\circ}=0$ the symmetry enhancement ceases at $O(z^{\prime 4})$ which is $\I_4^{\rm s}$ and the gauge group is $SU(4)$.

To see the $SU(2)$ part, we change the reference surface. The factors in (\ref{smsurface}) suggest
\begin{equation} \textstyle
 z'' \equiv  z' + a_1 b_5.
\end{equation}
The parameters become
\begin{equation}
 \begin{split} \label{smsurface2}
\a_1 &= -(b_5 + b_4 a_1) + O(z''),\\
\a_2 &= (b_4 + b_3 a_1)(z'' -a_1 b_5) + O(z^{\prime \prime 2}),\\
\a_3 &= -(b_3 + b_2 a_1)(z''- a_1 b_5)z'' + O(z^{\prime \prime 3}), \\
\a_4 &= (b_2 + b_1 a_1)(z''- a_1 b_5 )^2 z'' + O(z^{\prime \prime 4}), \\
\a_6 &= b_0 ( z''- a_1 b_5 )^3 z^{\prime \prime 2} + O(z^{\prime \prime 6}),
 \end{split}
\end{equation}
so that ${\rm
  deg}(\a_1,\a_2,\a_3,\a_4,\a_6,\Delta)=(0,0,1,1,2,2)$ at $\{z''=0\}$. From Table \ref{t:tate}, we check it is $\I_2$ for $SU(2)$.
 The discriminant has the form 
\begin{equation}
 \Delta =  \left( (b_5-a_1 b_4)^2 - 4 a_1^2 b_3 b_5 \right)^2 P_X^3
 P_{q_\circ}^3 P_{l_\circ} z^{\prime 2} +  P_{q_\circ}^2 P_X^2  P'
 z^{\prime 3} + P_{q_\circ} P_X Q' z^{\prime 4} + O(z^{\prime 5}).
\end{equation}
Again we used the same parameters in Table \ref{t:mequation}, and
$P'$ and $Q'$ are defined by this equation, which are not proportional
to $a_1$ or $b_5$. We see also vanishing of either
$P_X$ or $P_{q_\circ}$ enhances the symmetry to $SU(5)$, whereas $P_l=0$ for
$SU(2)$ charged lepton doublets $l$ enhances the symmetries to
$SU(3)$.

\subsubsection*{Perturbative limit and $U(1)_Y$}

Why our discriminant (\ref{SMdiscr}) is not of the factorized form, as familiar in $\I_5^{\rm s} \to \I_3^{\rm s} \oplus \I_2$ decomposition \cite{KV}?
It is because the {\em embedding group is $E_8$}, not $SU(5)$, otherwise we could not have desired matter contents but $X$-boson, from the mechanism (\ref{KatzVafa}).
Nevertheless we can see the factorization structure.
In the weakly coupling limit $\tau \to i\infty$, or large $f,g$ and finite $f^3/g^2$, F-theory can be approximated to type IIB string. In this picture, our symmetry is described by D-branes only (or one kind of $(p,q)$-branes) thus we see such separation and recombination of brane stacks. Specifically, for
small $a_1$, the discriminant (\ref{SMdiscr}) looks
\begin{equation} \label{A5toA2A3} \begin{split}
 P_{q_\circ} & \simeq P_{u^c_\circ} \simeq P_{e^c_\circ} \simeq b_5, \quad P_{l_\circ} \simeq P_{d^c_\circ} \simeq R_{\bf 5}, \\
 \Delta & \simeq b_5^4 R_{\bf 5} (b_5 a_1 + z')^2 z^{\prime 3},
\end{split}
\end{equation}
where $\{b_5=0\}$ and $\{R_{\bf 5}=0\}$ localize respectively $\{q_\circ,u^c_\circ,e^c_\circ\} \in {\bf 10}$ and $\{ d^c_\circ,l_\circ \} \in {\bf \overline 5}$.
This is analogous to separation of D-branes, but here $a_1 b_5$ is not a constant but a section of $-t$, thus the brane stacks intersect at angle.
In fact, the structure of $\P^1=\P({\cal O}_S \oplus -t)$ fibration is, as in Hirzebruch surface, linearly equivalent divisors are not necessary parallel but have self-intersections $\sigma \cap \sigma \ne 0$ as in (\ref{disjoint}).
Since the divisors $S = S_{SU(5)} = \{z=0\}$, $S_{SU(3)}=\{z'=z+\frac35 a_1 b_5=0\}$ and $S_{SU(2)}=\{z''=z-\frac25 a_1 b_5=0 \}$ are linely equivalent, we may use the notation $S$ without reference to gauge group if not necessary.
In the sense that the group is obtained by usual deformation, there is no problem in regarding the $S$ supports the single semisimple group.
Even in the perturbative limit,
we may take intersecting branes form a unified, connected cycles \cite{Choi:2006hm}.
Specifically, the discriminant components for $SU(3)$ and $SU(2)$ are linearly equivalent, so the internal cycles that worldvolumes wrap are at least homologous.

As expected, $U(1)_Y$ symmetry is the relative center-of-mass motion of the brane stacks, so it is reflected as the distance between the two proportional to $a_1$. We may require vanishing the trace part of $S[U(5)_\bot \times U(1)_Y]$. The choice
\begin{equation*} \textstyle
 z' \equiv z -\frac25 a_1 b_5 \ \Longrightarrow \ \Delta \simeq  b_5^4 R_5 (z+\frac35 a_1 b_5)^2(z-\frac25 a_1 b_5)^3,
\end{equation*}
fixes the center of mass at of the entire branes at $z=0$, making the total $SU(5)$ traceless, i.e. no $z^4$ term.
We have a hint for the existence of $U(1)_Y$ that the matter curves for $X$ and $q_\circ$ are distinctive, however identifying it as gauge symmetry with massless boson is not a trivial issue \cite{Hayashi:2010zp,Grimm:2010ez,Marsano:2010ix}.
By assumption of K3 fibration, the normal space is well-defined $\P^1$ and we can use $z$ as its global affine coordinate.
Although we have been interested in the properties of parameters on $S$, definitely the parameters $z,a_1,b_n$ are respectively sections of  $N_{S/B},K_B,K_B^{n-6}\otimes N_{S/B}$. 
So we have a {\em global} description of singularity that is exact to $O(z^5)$.
Since we will not turn on flux on this $U(1)_Y$ direction, we have neither induced Fayat--Illiopoulos term nor anomaly in six and four dimensions.

This shows also the unique position of the Standard Model group inside $E_8$, as $E_3 \times U(1)_Y$. We can easily see that other III or IV singularities in Table \ref{t:tate}, corresponding to $SU(3)$ or $SU(2)$, cannot have $\I_5$ or $SU(5)$ enhancement direction, hence no unification relation. This means we cannot guarantee the desired spectrum from the branching of $E_8$ for III or IV curves.

This model is consistent with gauge coupling unification, in the top-down sense.
The four dimensional gauge coupling is inversely proportional to volume that the supporting surface of the gauge theory wraps. Taking $SU(5)$ limit by $a_1 \to 0$, the volume of $SU(3)$ and $SU(2)$ are the same. (In some class of Calabi--Yau manifold, homologous supersymmetric cycles are calibrated, giving rise to the same effective volume for arbitrary $a_1$.) Also in this case, there is no contribution from $U(1)_Y$ flux required from the breaking of $SU(5)$ \cite{Blumenhagen:2008aw}.

\subsection{Associated matter curves} \label{s:mcurve41}

We turn to calculate nontrivial matter curves, for fields having antisymmetric tensor representations under the structure group. Already, using group theory,
we know their defining equations in Table \ref{t:mequation} as polynomials in $a_i,b_i$. 
For later calculation on the number of generations, we need to know 
the full homology of the matter curves on the spectral cover.
For this, we derive the matter curves from the combinations of spectral covers, using the equations of $F_X,F_5$ in (\ref{su5xu1eq}).

\subsubsection*{The $d^c_\circ$ curve} \label{ss:d}

The field $d^c_\circ$ ${(\bf 5,10)}$, shown in Table \ref{t:mcurve}, transforms as the rank two tensor representation ${\bf 10}$ under $SU(5)_\bot$, thus is localized on the corresponding matter curve
\begin{equation} \label{inducedrel}
 t_i + t_j = 0,  \quad 1 \le i < j \le 5.
\end{equation}
These are to be related to the curve $t_i=0$, as the tensor structure gives a combination of the corresponding weights. Therefore we want to obtain the matter curve as an induced object from the fundamental one $C_5$, which is called as associated matter curve.
Since each weight vector is related to a line bundle, the condition (\ref{inducedrel}) is a relation between two line bundles ${\cal O}(t_j) = {\cal O}^{-1}(t_i)= {\cal O}(-t_i)$.
Thus the solution comes from the common intersection $F_5(t_j)=F_5(-t_i)$
for every $t_i$. Thus our matter curve is contained in \cite{FMW}
\begin{equation} \label{d0curve}
 F_5(t) = F_5(-t) = 0.
\end{equation}

The flipping of the signs of all the roots
$t_i,i=1,\dots 6$ is realized by an involution $\tau$
\begin{equation} \label{involution}
 \tau: V \leftrightarrow -V.
\end{equation}
It becomes the orientifold projection in the type IIB theory limit.
We may say the cover $C_5$ is fundamental, since the others
follow from the common intersections. So,
(\ref{inducedrel}) is an induced component on $S$  inside the common
intersection
\begin{equation} \label{CcapC}
 \tau C_5  \cap C_5 .
\end{equation}
However we should drop some redundant parts, as follows. The relation (\ref{CcapC}) does not care about $i<j$ condition. Setting $i=j$ in (\ref{inducedrel}) gives $t_i=0$, which gives already known the matter curve for $q_\circ$, namely $C_5 \cap \sigma$.
Also non-restriction of $i<j$ means that (\ref{CcapC}) contains two identical copies or double cover of the matter curve.

The common intersection of (\ref{d0curve}) is equivalently
obtained as the intersection of $F_5(t) \pm F_5(-t)=0$. That is
\begin{equation} \label{Vevenodd}
\begin{split}
 U (b_0 U^4 + b_2 U^2 V^2 + b_4  V^4) = 0, \\
 V ( b_1 U^4  + b_3 U^2 V^2 + b_5 V^4) = 0.
 \end{split}
\end{equation}
The solution to these in general would give a fourfold cover, but
we know we should have a double cover of the matter curve.
 So, we eliminate $b_0 U^5$ and $b_1 U^4 V$
terms to have
\begin{equation} \label{V2factor}
 \begin{split}
0 &= V^3  \big( (b_1 b_2 - b_0 b_3) U^2 + (b_1 b_4-b_0 b_5) V^2 \big)  \\
 & =-V^3 a_0^{-1} \big( (a_1b_0 b_2 +a_0 b_0 b_3) U^2 + (a_1 b_0 b_4+a_0 b_0 b_5) V^2 \big).
\end{split}
\end{equation}
We keep in mind that $b_1$ is a derived quantity (\ref{su6traceless}).

All of the solution spaces are linearly equivalent to $[C_5]$.
An expected solutions is $U=b_5=0$, corresponding to the
matter curve $\Sigma_{q_\circ}$. $V=b_0=0$ corresponds to a threefold intersection at infinity that we cannot access from $S$, so we drop it.
From (\ref{V2factor}) we plug back
\begin{equation} \label{twosheets}
U = \pm i \sqrt{\frac{b_1
    b_4-b_0b_5}{b_1 b_2-b_0 b_3}}V = \pm i \sqrt{\frac{a_1 b_4+a_0
    b_5 }{a_1  b_2+a_0  b_3}}V,
\end{equation}
to obtain
\begin{equation} \label{coverd}
  -b_0^{-1} V^4 (a_1  b_2+ a_0 b_3)^{-2} P_{d^c} = 0.
\end{equation}
Thus we find the matter curve $P_{d^c_\circ} = 0$ for $d^c_\circ$, shown in Table
\ref{t:mequation}.
We can identify each term, so does the matter curve, 
transforms as section
of $3\eta-10c_1+2y$ on $S$, displayed in the third column in Table
\ref{t:mcurve}. For $y=0$ it reduces to $\bf 5$ curve (\ref{R5}), showing the inheritance relation from $SU(5)$ GUT.
Two possible solutions in (\ref{twosheets}) correspond two covers, but
they just give the same matter curve.

We calculate the corresponding homology class. The above equations
show that there are redundant components in $C_5 \cap \tau
C_5$. Since we dropped
$U$ and $V^3/a_0$, homologically we have
\begin{equation} \label{cd}
\begin{split}
 [{\cal P}_{ d^c_\circ}] =  ([C_5] &-\sigma) \cap ([C_5]-(3\sigma_\infty-\pi^*y)) \\
 &  = (4\sigma+\pi^*\eta) \cap (2\sigma+\pi^*(\eta-3c_1+y)) \\
 &= 2 \sigma \cap \pi^*(3\eta-10c_1+2y)+ \pi^*\eta \cap \pi^* (\eta-3c_1+y).
  \end{split}
\end{equation}
The first term is linear in $\sigma$, meaning that the corresponding
component is on $S$, which is usually called as $\Sigma_{d^c_\circ}$.
The second term lies outside $S$ but only on $C_5$.
In this
sense, the spectral cover is more fundamental object containing localization
information on matters. The factor
$2$ in front displays that it is double cover of the (expected) matter
curve $3 \eta -10 c_1 +2y$, shown in Table \ref{t:mcurve}.
In what follows, we will omit pullback $\pi^*$ and denote the object itself also as homological cycle without confusion.

As in the derivation of $q_\circ$ and $X$ curves in (\ref{fundmattercurves}),
we may attempt to build the corresponding spectral cover $C_{\wedge^2 V} \sim \prod_{i<j}(s-t_i-t_j)$ from which $C_{\wedge^2 V} \cap \sigma$ would induce the matter curve $d^c_\circ$.
It is known that, however, it has severe singularities at $t_i + t_j
= t_k + t_l$ where all the indices are different, and the corresponding spectral line bundle is not found in $\v Z$. Although this is overcome by normalizing the curve \cite{Donagi:2004ia,Hayashi:2008ba}, we will take the same result from a slightly different interpretation.

\subsubsection*{Other curves}

Other matter curves follow straightforwardly.
For $u^c_\circ$, we take the intersection $C_5 \cap \tau C_X$. In equation we have
\begin{equation} \label{uccurve} \begin{split}
 0 &= V^5 a_0^{-5} \left( b_0 a_1^5+b_1 a_0 a_1^4 + b_2 a_0^2 a_1^3
+ b_3 a_0^3 a_1^2+ b_4 a_0^4 a_1 + b_5 a_0^5 \right) \\
  &= V^5 a_0^{-3} P_{u^c_\circ}.
  \end{split}
\end{equation}
Here the first two terms cancel in the first line due to tracelessness
(\ref{su6traceless}), resulting in the matter curve equation $P_{u^c_\circ}$ in Table \ref{t:mequation}. From each coefficient, we expect it transforms on $S$ as
$\eta-5c_1+3y$. We have
another nontrivial solution of multiplicity two
$V=a_0=0$ in (\ref{uccurve}), so that
\begin{equation}
 {\cal P}_{ u^c_\circ}=
  C_5  \cap  C_X -2 \sigma_\infty \cap y
 =  (C_5 - 2 \sigma_\infty) \cap C_X
  =  \sigma(\eta-5c_1+3y)+(\eta-2c_1) y.
\end{equation}
There is no more common solution, such as $U=a_1=0$ since
$b_5$ is not automatically zero. As in the $SU(5)$ case, though
seemingly different, we cannot distinguish this from $\tau C_5
\cap C_X$, so it is not independently counted.

As a consistency check, the total sum of the curves including the
intersections at infinity becomes
$$ (C_5 \cup C_X) \cap \tau (C_5 \cup C_X) = {\cal P}_{d^c_\circ} \cup {\cal P}_{u^c_\circ} \cup {\cal P}_{q_\circ} \cup {\cal P}_{X} \cup (C_5 \cap 3 \sigma_\infty) \cup (C_X \cap 3
\sigma_\infty).
$$


The matter curve for $e^c_\circ$ comes from the intersection between $C_5$
and $C_X$, calculated in the same manner as $u^c_\circ$ without any involution. There is no
cancellation from $b_0 a_1^5 -b_1 a_0 a_1^4 = - b_1
a_0 a_1^4$, thus we obtain $V^5 a_0^{-4} P_{e^c} =0 $ in Table \ref{t:mequation}.
The corresponding homology class is
\begin{equation} \label{covere}
 {\cal P}_{ e^c_\circ}
 = (C_5 - \sigma_\infty) \cap C_X
 = \sigma(\eta-5c_1+4y) +(\eta- c_1) y
\end{equation}
which is to be distinguished from ${\cal P}_{u^c}$. The quantum number is tracked back to the adjoint of $SU(6)_\bot \to SU(5)_\bot \times U(1)_Y$ so the corresponding field is inherited from the moduli. For pure moduli living on $C_5$, we cannot calculate the homology class \cite{FMW,DW3}. We name the remaining moduli as $\nu_M^c,M=1,2$ which belong to
\begin{equation}
 h^{2,0}(C_5) \oplus h^{0,1} (C_5).
\end{equation}
Thus we can know the number of singlet neutrinos after specifying the Calabi--Yau manifold.

For $l_\circ$, which is not yet distinguished from $h_{d\circ},
h_{u\circ}^c$, we need the product
$ \prod(t_i+t_j+t_6)$ with $1 \le i < j \le 5$. This of course gives
the matter curve $P_{l_\circ}$ in terms of $a_m,b_m$ in Table \ref{t:mequation}.
This corresponds to the rank three antisymmetric representation $\bf
20$ of $SU(6)$. The constraint on the line bundle is ${\cal O}(t_k) = {\cal O}^{-1}(t_i+t_j) = {\cal O}(-t_i) \otimes {\cal O}(-t_j)$. Thus, we expect that it is contained in  the common intersections of
$$  F(t_k)= F(-t_i-t_j)=0, $$
lying on a codimension one curve.
In fact the first equation also implies $F(t_k)=F(t_i)=F(t_j)=0$, we can solve the equation.
In practice it is not necessary to calculate its explicit homology since the consistency requires the contribution from $t_6$ is trivial. In the limit of $SU(5)$ the matter
curve should reduce to $\bf \overline 5$ of $SU(5)$ which is identical to that
of $d^c_\circ$. This should be, if there is $SU(5)$ relation connecting fields in $\bf \overline 5$.

\subsubsection*{Green--Schwarz relation in six dimensions}

The anomaly freedom of a compactified F-theory is taken care of by the Green--Schwarz (GS) mechanism, generalizing that of type IIB theory.
The theory so far is six dimensional.
There, due to the simplicity of GS polynomial,
the relations restrict the possible homology classes of charged matter curves
\cite{Sadov:1996zm,Choi}. From these we can extract the relation between anomaly coefficients and group invariants by observing local divisors supporting specific gauge theory \cite{Sadov:1996zm}. We quote some of relevant results. For all the matter curves $[\Sigma]$ on each supporting surface $S_i,i=SU(3),SU(2),$ we have
\begin{align}
 \ell({\rm adj}_i) - \sum_R \ell(R_i)  [\Sigma_{R_i}] &= 6 K_S \cdot S_i, \\
 \sum_{R,R'} \ell(R_i) \ell(R_j') [\Sigma_{R_i,R_j'}] &= S_i \cdot S_j.
\end{align}
Here $\ell (R)$ is the index of the representation $R$, or $\tr_R t_a t_b = \ell(R) \delta_{ab}$. We displayed only relevant quantum numbers.
Applying the relations for $SU(3)$ and $SU(2)$, we obtain
\begin{equation} \label{6dgsrel} \begin{split}
 \sum_{R \in SU(3),R'}  \ell(R){\rm dim}(R') [\Sigma_{R,R'}] &=  9c_1 - 6t, \\
 \sum_{R \in SU(2),R'} \ell(R) {\rm dim}(R') [\Sigma_{R,R'}] &= 8c_1 - 6t, \\
 \sum_{R \in SU(3),R' \in SU(2)} \ell(R)\ell(R') [\Sigma_{R,R'}] &= -t.
  \end{split}
\end{equation}

From (\ref {fundmattercurves}), the last condition particularly implies on $q_\circ$ and $X$, both of which are $\bf (3,2)$
\begin{equation}
  [\Sigma_{q_\circ}] + [\Sigma_X] = (\eta-5c_1) + ( -c_1 + y)= -t =  \eta-6c_1,
\end{equation}
with $\ell({\bf 3})=\ell({\bf 2})=1$.
So this shows $y=0$, that is, $a_0$ should be a trivial bundle. The result seems to indicate that the choice of the homology for $a_0$ hence $a_1$ cannot be arbitrary.
The triviality of $a_0$ comes from the fact that we recycled the parameters $b_i$ as those of $SU(5)$. Since the Calabi--Yau condition states that the combination $a_0 b_0$ should be a section of $\eta-6c_1$, we have no extra degree of freedom to choose $a_0$.
In six dimensions, all the gauge theory appearing in F-theory is understood as chains of higgsing from a larger gauge symmetry \cite{Beetal}, which does not change the above GS relations after symmetry breaking.
Since we can obtain the SM group from the $SU(5)$ GUT group formally by higgsing of ${\bf 24}$, we inherit the parameters $b_i$ of $SU(5)$. In Ref. \cite{Beetal}, for instance, $SO(10)$ parameter $q_{6+n}$ is inherited from $E_6$ curve, otherwise two parameters are polynomials of different degrees.

We can check that, then, the other relations are automatically satisfied and each six dimensional effective gauge theory has no anomaly. So in what
follows we take $a_0=1$.

\subsection{Matter curves under nontrivial monodromy}

\begin{table}[t]
\begin{center} \begin{tabular}{cc}
\hline  section & transformation \\
\hline
$U$ & $\sigma$ \\
$V$ & $\sigma+\pi^*(c_1)$ \\
$a_k$ & $\pi^*(y-kc_1)$ \\
$d_k$ & $\pi^*(x-kc_1)$ \\
$e_k$ & $\pi^*(\eta-kc_1-x)$ \\
\hline
\end{tabular} \end{center}
\caption{Transformations of the coefficients of (\ref{factor411}). $x$
and $y$ are arbitrary linear combinations of $\eta$ and $c_1$.}
\label{t:transf}
\end{table}

When we count a matter field, we considered its identity by the orbit
of matter curves under a given monodromy group.
The spectral cover should
encode such information, since the matter curves are induced from it.
By using the parameters $b_0,b_1,\dots b_5$,
so far we have implicitly assumed $S_5$ monodromy. As seen in Sec. \ref{sec:z4monodromy} we need $\Z_4$ monodromy for the realistic spectrum so here we study how to realize its matter curve.

\subsubsection*{The $\Z_4$ monodromy}

The spectral cover in fact describes a freedom of {\em choosing} monodromy of the background gauge bundle. We go on with $\Z_4$ monodromy
considerd in Section \ref{sec:z4monodromy}.

Since $\Z_4$ singles one cover out of five, accordingly
the spectral cover (\ref{su5xu1cover}) is further factorized as
$C_X \cup C_5 \to C_X \cup C_{q'} \cup C_{q}$,
described by
\begin{equation} \label{factor411}
(U +a_1V)
(d_0  U + d_1 V)
(e_0 U^4 + e_1 U^3 V + e_2 U^2 V^2 + e_3 U V^3 + e_4 V^4) = 0.
\end{equation}
This is understood as a tuning among the parameters of
$C_5$
\begin{equation} \begin{split}
 b_0 &= d_0 e_0,\\
 b_i &= d_0 e_i + d_1 e_{i-1},\quad i=1,\dots,4, \\
 b_5 &= d_1 e_4.
  \end{split}
\end{equation}
(We can understand the embeddability of $SU(5)_\bot \times U(1) \subset
SU(6)_\bot$ for (\ref{su5xu1cover}) in the same way.)
The traceless condition (\ref{su6traceless}) becomes
\begin{equation} \label{finaltraceless}
 a_1 d_0 e_0 + d_1 e_0 + d_0 e_1 = 0.
\end{equation}
Again we have an overall freedom to choose $d_0$, whose transform we name $x
\in H_2(S,\Z)$. The transformations of parameters are summarized in Table \ref{t:transf}.
Relating $d_1/d_0 \sim  t_5$, it is easy to see that $e_i/e_0$ are
again the elementary symmetry polynomials of degree $i$ of
$t_1,t_2,t_3,t_4$, related by $S_4$ monodromy subgroup.
The extra $U(1)_M$ in (\ref{u1bminusl}) is now identified by the one generated by cover $C_{q'}$, as to what $SU(1)_Y$ is by $C_X$.

As expected, we obtain `fundamental' matter curves for $X,q'$ and $q$ as
\begin{equation} \label{cover}
 \begin{split}
 {\cal P}_X & = C_X \cap \sigma = -c_1 \cap \sigma \\
 {\cal P}_q  & = C_q \cap \sigma =  (\eta-4c_1-x) \cap \sigma \\
 {\cal P}_{q'}&= C_{q'} \cap \sigma =  (-c_1+x) \cap \sigma
 \end{split}
\end{equation}
where we omitted pullback.

Second, we decompose $e_i$ into irreducible ones under the
monodromy $\Z_4$. This is done by further tuning $e_i$'s. From Table \ref{t:mcurve41}, we note that only $\{ t_i + t_j \}_{1\le i<j\le4}$ have nontrivial representations compared to that of $S_4$. For this, we can show that the only needed decomposition is
\begin{equation} \label{e2decomp}
 e_2 = e_2' + e_2''
\end{equation}
making closed $\Z_4$ orbits
\begin{equation}
 e_2'/e_0 \sim  t_1t_2+t_2t_3+t_3t_4+t_4t_1, \quad e_2''/e_0 \sim
  t_1t_3+t_2 t_4.
\end{equation}
It follows that the matter curve inherited from $d_\circ^c$ in Sec \ref{ss:d}
\begin{equation} \label{S4closed}
 \prod_{1 \le i< j \le 4} (t_i +t_j) \sim e_0^{-3} (-e_0 e_3^2+e_1 e_2
  e_3-e_1^2 e_4),
\end{equation}
containing six factors, is factorized into $\Z_4$ irreducible ones
\begin{equation} \label{Z4irred} \begin{split}
 D_1^c&: (t_1+t_3)(t_2+t_4) \sim e_2'/e_0, \\
 D_2 &: (t_1+t_2)(t_2+t_3)(t_3+t_4)(t_4+t_1) \sim (e_2^{\prime
   \prime 2}+e_1e_3-4e_0e_4)/e_0^2.
  \end{split}
\end{equation}
These relations are obtained using group theory, and decomposition of (\ref{S4closed}) to (\ref{Z4irred}) {\em defines} the parameters
$e_2'$ and $e_2''$ in (\ref{e2decomp}).
In other words, this factorization structure of coefficients in the spectral cover equation describes $\Z_4$ monodromy (See also \cite{Marsano:2009gv}).

On $S$, the homology classes for $D_1^c$ is $\eta-2c_1-x$. Noting that $e_1$ should be expressed in terms of $e_0$ in (\ref{finaltraceless}), the corresponding class is $(3\eta-6c_1-3x)-(\eta-x)=2\eta-6c_1-2x.$
Since $D_2$ is the remaining part of the decomposition, we implicitly had a nontrivial condition that $e_2''$ has a factor $e_0$, which gives the overall class as $(2 \eta-4c_1-2x)-(\eta-x) =\eta-4c_1-x$. The results are displayed in Table \ref{t:mcurve41}.

\subsubsection*{Associated matter curves}

We need the homology class $\cal P$ inside $\v Z$ for each associated matter curve.  Since $y$ distinguished the homology of $u^c$ and $e^c$, so the above setting $y=0$ makes various curves
homologous. In particular it becomes $C_X = \sigma$ and it follows
\begin{equation} \begin{split}
 {\cal P}_{u^c} = & \  \tau (C_q-2\sigma_\infty)\cap C_X = (\eta-4c_1-x)\cap \sigma \\
  &= (C_q-\sigma_\infty)\cap C_X = {\cal P}_{e^c},
  \end{split}
\end{equation}
where we used the disjoint relation (\ref{disjoint}).
Of course, the actual equations for $u^c$ and $e^c$ are different, as seen in Table \ref{t:mequation}, so the corresponding matters are localized on different curves. They are homologous. Likewise
$u'$ and $e'$ are respectively distinguished by involution $\tau$
but homologous
\begin{equation} \begin{split}
 {\cal P}_{u'} =&\  C_{q'}\cap C_X  =  (-c_1+x)\cap \sigma \\
  & = \tau C_{q'} \cap C_X= {\cal P}_{e'}.
  \end{split}
\end{equation}
These relations imply $SU(5)$ GUT unification structure where $\{q,u^c,e^c \}\in {\bf 10}$, as a consequence of the triviality of $a_0$.

We can perform a similar calculation for $d^c_\circ$ and $l_\circ$.
Since those for $d^c$ and $l$ are equivalent in trivial $a_0$ or $t_6 \to 0$, we continue the $d^c$ part.
The matter curve for $d^c$ is inside $\tau C_{q'} \cap C_{
  q}$. The calculation is analogous to the case of $u^c,e^c$ curves,
\begin{equation}
 V^3 d_0^{-3} (-a_1 d_1^3 e_0 +  e_2  d_0d_1^2 + e_3 d_0^2 d_1 + e_4 d_0^3).
\end{equation}
Thus we have
\begin{equation}
 {\cal P}_{d^c} = {\cal P}_l  = \tau C_{q'} \cap (C_q-\sigma_\infty)=
 \sigma(\eta-4c_1+2x)+x(\eta-c_1-x).
\end{equation}
The matter curves for $D_1$ and $D_2$ are obtained by decomposing
$C_q \cap \tau C_q$ referring to the homology class displayed in Table \ref{t:mcurve41}.
The corresponding classes are
\begin{equation} 
 {\cal P}_{D_2}= {\cal P}_{h_u} =\  (C_q-2\sigma) \cap (C_q-4\sigma_\infty)
   = 2 \sigma(\eta-4c_1-x)+ (\eta-x)(\eta-4c_1-x) \\
\end{equation}
and 
\begin{equation}
 {\cal P}_{D_1^c}= {\cal P}_{h_d^c} =  C_q \cap (\sigma+\sigma_\infty) =
2 \sigma(\eta-2c_1-x)+ c_1(\eta-x) 
\end{equation}
which is also equivalent to the cycles for $h_d^c$ and $\nu_1^c$.
We distinguish the physical lepton doublets from Higgs doublets.

To agree with the previous matter curves before $\Z_4$ modding, the
total sum of homology class in the unified multiplet should be the
same. We have
\begin{equation} \begin{split}
 {\cal P}_{u^c_\circ} &= {\cal P}_{u^c} \cup {\cal P}_{u'} \\
 {\cal P}_{e^c_\circ} &= {\cal P}_{e^c} \cup {\cal P}_{e'} \\
 {\cal P}_{d^c_\circ} & =  {\cal P}_{d^c} \cup {\cal P}_{d} \cup {\cal P}_{D_1^c}
 \cup {\cal P}_{D_2} \cup \tau(C_{q'}-\sigma) \cap (C_{q'}-\sigma_\infty), \\
 {\cal P}_{l_\circ} & =  {\cal P}_{l} \cup {\cal P}_{l^c} \cup {\cal P}_{h_u^c}
 \cup {\cal P}_{h_d} \cup \tau(C_{q'}-\sigma) \cap (C_{q'}-\sigma_\infty), \\
  \end{split}
\end{equation}
where the last terms in the last two lines will vanish later in $x\to 0$ limit. As before, we
cannot distinguish ${\cal P}_{d^c} = \tau (C_q-\sigma_\infty) \cap C_q'$ from ${\cal P}_{d} \equiv ( C_q
-\sigma_\infty) \cap \tau C_q'$, so we count only one of them. This provides
another understanding on $SU(5)$ unification structure of the matter fields.

\section{Four dimensional spectrum}

So far, we have identified matter contents localized along
curves in the internal manifold. In total space including our four dimensions, each curve spans six dimensional worldvolume, thus its spectrum is yet non-chiral in four dimension.
In order to obtain chiral spectrum, we should turn on magnetic flux $F$ on the matter curves \cite{Magflux}. 
We use a line bundle $\cal N$ in Cartan subalgebra (\ref{Cartans}), or in terms of field strength $G = \sum F^i \wedge \omega$, so we call this as $G$-flux \cite{FMW}.
The supersymmetry conditions for these are Hermitian Yang--Mills equations
\begin{align}
 \text{F-flat}&: F^{(2,0)} = F^{(0,2)}=0, \\
 \label{Dflat}
 \text{D-flat}&: J \wedge F + \frac i2 [\varphi,\varphi^\dagger] =0,
\end{align}
with the K\"ahler form $J$ on $S$.

Although physically different, the spectral cover describing
the vector bundle background looks like another stack of seven-branes, whose positions from some origin are nothing but the eigenvalues $t_i$, that is, along the fiber $E$ the background vector bundle is decomposed,
$$ {\cal V}|_E = \bigoplus {\cal O} (t_i). $$
Turning on a $G$-flux, or spectral line bundle $\cal N$, on the spectral
cover $C$ induces the flux on the matter curves.
Being on
$C$, it {\em does not} break the gauge symmetry living on $S$. The
zero modes of four dimensional chiral fermions are classified by
homology, whose unified description is done by Ext group \cite{Distler:2007av,Hayashi:2009ge,DW3}. Letting $i,j$ respectively immersions of
$S$ and $C$ into $\v Z$ in (\ref{threefold})
\begin{equation} \label{ascflux}
 {\rm Ext}^1 (i_*{\cal O} , j_* {\N}) = H^0\big(\Sigma,({\N} +
 K_S)_\Sigma \big),
\end{equation}
classifying the Dirac eigenstates along the complex curve $\Sigma \equiv C \cap \sigma$ as
topological defect \cite{BHV1}.

We can regard the above as intersecting brane picture.
Since F-theory employs open stringy description, where branes are natural objects for specifying the boundary conditions, they provide two distinctive way of obtaining chiral gauge theory in four dimensions. 
GUT realizations are regarded as magnetized brane description: turning on flux
$\cal L$ of the field strength for the gauge group on $S$.
For example the $U(1)_Y$ flux achieve the breaking of the GUT group and realize chiral four dimensional spectrum. It always
breaks the gauge symmetry on $S$, its gaugino field decomposes and gives
rise to chiral fermions in four dimension, belonging to
$$ {\rm Ext}^2 (i_*{\cal O}, i_* {\cal L}) = H^1(S,{\cal L} +
K_S) \oplus H^0 (S,{\cal L}). $$
Since we start with the Standard Model group from the beginning, we do not need this mechanism. 

\subsection{Flux on a spectral cover}

We turn on $G$-fluxes on the spectral cover, which hence induce fluxes on
the matter curves on it. Since the spectral cover is factorized as in
(\ref{factor411}), we can turn on the flux differently on each cover, provided that consistency conditions are satisfied.

Let $C$ be a $n$-fold spectral cover of $S$ and $p_C$ be its projection
\begin{equation}
 [C]=n \sigma + \pi^*\eta,\quad p_{C}: C \to S.
\end{equation}
As alluded, for four dimensional chiral spectrum, we turn on a line bundle $\N$ on $C$
\begin{equation} \label{intN}
 c_1(\N) \in H^2(C,\Z),
\end{equation}
which is related to a rank $n$ vector bundle $\V$ of structure group $U(n)$ (of the Higgs bundle or, of the vector bundle in the heterotic dual). This correspondence is known as Fourier--Mukai transformation \cite{FMW}. Their
 characteristic classes are related by the Grothendieck--Riemann--Roch
index theorem \cite{FMW}. The first Chern classes are related as
\begin{equation} \label{traceless}
 c_1(\V)=p_{C*} \left( c_1({\cal N})+\frac12(c_1(C)-p_C^* c_1) \right).
\end{equation}
The pullback $p_C^*$ is $n$-fold lifting and the
pushforward $p_{C*}$ loses such information, we have
$p_{C*} p_C^* =n$. 
If we want $SU(n)$ rather than $U(n)$, we may make it traceless by imposing
$c_1(\V)=0$. Then, we have another equivalent
\begin{equation} \label{trlesflux}
 c_1 ({\cal N})=\frac12 (-c_1(C)+p_C^* c_1)+  \lambda \gamma,
\end{equation}
for a class $\gamma \in C$ satisfying
\begin{equation} \label{tracelessproj}
 p_{C*} \gamma = 0.
\end{equation}
Here $\gamma$ is the one we specify, so we call also this as $G$-flux.

A rational number $\lambda$ is
introduced to make $\N$ be in the integral class (\ref{intN}).
The quantity in the parenthesis of (\ref{traceless})
is called the ramification divisor, reflecting the multifold nature of $\cal N$ in $\V$.
We constructed $C$ as a subspace of threefold $\v Z$ in
(\ref{threefold}). In this case,
using the adjunction formula for $c_1(\v Z)$ on $C$,
\begin{equation}
 -c_1(C)+p_C^* c_1=\left(-c_1(\v Z)+C \right)_C+p_C^* c_1 = (n-2) \sigma + p_C^*(\eta - c_1),
\end{equation}
the relation (\ref{trlesflux}) becomes explicit
\begin{equation}
 c_1(\N) = \textstyle (\frac{n-2}{2} +n\lambda) \sigma +  (\frac12-\lambda)\pi^* \eta +
 (n\lambda-\frac12)\pi^* c_1.
\end{equation}
As a result the integral condition (\ref{intN}) becomes
\begin{equation} \textstyle
 n (\frac12+\lambda) \in \Z, \quad (\frac12-\lambda)\pi^*\eta+(n\lambda-\frac12)\pi^* c_1 \in
H^2(S,\Z).
\end{equation}
As we see in the next section, a large class of base manifolds admits
$\eta-c_1$ as even cycles. So we just require $\lambda$ to be
any integer for even $n$, or a half-odd-integer for odd $n$. Turning off the flux $\lambda=0$ requires even $n$.
There is no such constraint for the $U(1)$ case, since there is no ramification
$c_1(C)-p_C^* c_1 = 0$.

Another necessary condition is D-flatness in (\ref{Dflat}).
Since the K\"ahler form $J$ is $(1,1)$ form, choosing $F \in H^{1,1}(S,\Z)$
satisfies the condition.
One obvious choice is a cocycle $\gamma$ dual to the matter curve $\Sigma \equiv C \cap \sigma$ \cite{FMW,DW1}.
Since $p_{C*} (\Sigma) = \eta - nc_1$, we have
\begin{equation}  \label{univflux}
 \gamma = (n - p_C^* p_{C *})(C \cap \sigma)
         = (n\sigma-\pi^*(\eta-nc_1)) \cap C.
\end{equation}
This is always present, hence called the universal flux.

Through (\ref{ascflux}) a line bundle $\N$ induces flux on a
matter curve  $\sigma$
as \cite{Curio}
\begin{equation} \label{spectrallinebundle} \begin{split}
 \left({\N} + K_S\right)_{\Sigma}
 &= \left(-\frac12 \Big(c_1(C)-p_{C}^* c_1\Big)+
  \lambda \gamma+ K_S\right)_{\Sigma} \\
 &= \frac12 \Big((\eta-nc_1) + \lambda \gamma \Big)_{\Sigma} \\
 &=  \frac12 K_{\Sigma} +
    \lambda \gamma|_{\Sigma}  .
  \end{split}
\end{equation}
Here in the second line we used $p_{C *}{\Sigma}=\eta-nc_1$ and in the last line the adjunction formula $K_\Sigma=(K_S+\Sigma)_\Sigma$.
Under this background, four dimensional massless chiral fermions are the zero modes of Dirac operator in $\Sigma$.
Their number difference is calculated by Riemann--Roch--Hirzebruch
index theorem
\begin{equation} \begin{split}
 n_f &\textstyle \equiv h^0 (\Sigma,{\N} + K_S) -h^1(\Sigma,
 {\N} + K_S )  \\
 &= \int_{\Sigma} {\rm Todd}(\Sigma) {\rm ch} ({\textstyle \frac12 K_\Sigma + \lambda \gamma}) \\
 &\equiv \int_\Sigma c_1({\textstyle \frac12 K_\Sigma + \lambda \gamma}) - \frac12 c_1(K_\Sigma) \\
 &= \lambda \int_\Sigma \gamma = -\lambda \eta \cdot (\eta-n c_1).
\end{split} \end{equation}
In the second line we used (\ref{spectrallinebundle}) and in the last line the contributions of $\frac12 K_\Sigma$ cancel.
Using Poincar\'e dual cycle with the same notation, we can express it as $\lambda \gamma \cdot \Sigma$.

Since associated matter curves $\cal P$, studied in Section \ref{s:mcontents}, are
derived from the intersection of the spectral cover, the flux on
the spectral cover should also induce a flux on the
associated curves \cite{Donagi:2004ia}. From the tensor structure of matter curves, this new flux  is derived from the line bundle $\cal N$ on $C$. For this generalization, let us define
$$\Gamma_C = n\sigma - p_C^* (\eta -n c_1).$$
From above we have seen $\Gamma \cap C = \gamma$, showing the dependence
on the spectral cover over which the flux is turned.
It is already seen traceless.
The index theorem for the associated matter curve shows
the spectrum is similarly obtained \cite{Donagi:2004ia},
\begin{equation} \label{Pmulti}
 n_f = \int \left(c_1(\N)+\frac12 c_1(K_S)+\frac14 c_1(\v Z) +\frac12 c_1({\cal
   P})|_{\cal P} - \frac12 R \right) = \lambda {\cal P} \cap \Gamma|_S
\end{equation}
Here the formula is essentially the same as before, where the contribution from $c_1(\v Z)$ was trivial. $\frac12 {\cal P}$ is the contribution from Todd class, which was $-\frac12  K_\Sigma$. The number of ramification points $R$ was counted to take the normalized matter curve. For example for $d^c_\circ$, we have
$$
 R =  {\cal P}_{d_0^c} \cap (\sigma+3 \sigma_\infty)= (C_5-\sigma-3\sigma_\infty) \cap C_5 \cap (\sigma+3\sigma_\infty).
$$
In the end, there are cancellation of terms except the $\Gamma$ dependent contribution in $\N$, as before.

The spectrum in the class $t_i +t_j + t_6$ gives rise to rank three antisymmetric tensor representation of $SU(6)$. Since we want unbroken $U(1)_Y$ untouched, it suffices to consider turning on flux
except $C_X$ cover corresponding to $t_6$. The resulting matter curve
is effectively antisymmetric curve $t_i+t_j$.

Some comments are in order.
We assumed that the matter curve comes from the same spectral cover component where the very flux $\N$ is turned on. Otherwise the flux cannot affect the chirality of the matter localized on another spectral cover component.
It is simply because the field is not charged under the corresponding dual structure group.
Also, Kodaira vanishing theorem on the one dimensional curve states that
$$ h^0(\Sigma,{\cal L}_\Sigma) = 0, \quad \deg {\cal L}_\Sigma \le 0, $$
and a similar for the antiparticle homology. Thus it is
nonzero only if the flux on the matter curve $\gamma \cap \Sigma$
has positive degree \cite{BHV1}. So it is difficult to obtain vectorlike components for the spectrum from the matter curve. For this reason we distinguish the matter curves for $h_u$ and $h_d$.

\subsection{Final spectrum using factorized flux}

We may consider turning on spectral fluxes (\ref{intN}) on various subsets of the covers
\begin{equation}
 \bigcup C_m, \quad m \in \{q,q',X\},
\end{equation}
which does not necessarily mean the union of all the covers.
The cover $C_X$ provides a new freedom that is not present in $SU(5)$ GUT models. We will shortly see that turning on fluxes on different part of covers leads to different unification relations among the matter contents.

The simplest choice is to turn on a flux only on $C_q$ \cite{CK}
\begin{equation} \label{su4flux}
 \Gamma_q   = (4-p_q^* (\eta-n c_1)), \quad
  \Gamma_{q'}=0, \quad \Gamma_X=0,
\end{equation}
satisfying a stronger version of (\ref{trlesflux}).
The matter spectrum is obtained by the above formula.
For example for $q$, we have
\begin{equation} \label{multipl}
 \begin{split}
 n_q   = \lambda {\cal P}_{q} \cap \Gamma_q &= \lambda (\eta-4c_1-x) \cap \left(4\sigma -\pi^*
 (\eta-4c_1)\right) \cap \sigma|_S\\
 & = \lambda (\eta-4c_1-x) \cap (-\eta) \cap \sigma|_S \\
 &= - \lambda \eta \cdot (\eta-4c_1-x).
 \end{split}
\end{equation}
We used the disjoint relation (\ref{disjoint}) and the inner product is done for divisors of $S$.
Being only on $C_q$, the flux is induced only on the components of associated matter curve parameterized by $t_i,i=1,\dots,4$. Thus we easily understand the same multiplicity for $u^c$ and $e^c$ as $q$
$$ n_{u^c} = n_{e^c} = n_q =  - \lambda \eta \cdot (\eta-4c_1-x).$$
The curve components of $t_5$ or $t_6$ are neutral, so that
\begin{equation}
 n_X= n_{q'}=n_{u'}=n_{e'}=0.
\end{equation}
Thus we do not worry about unwanted coupling involving these fields, as shown in Table \ref{t:Yukawa}. This is to be contrasted to the case of the $SU(5)$ GUT, broken by a nontrivial flux along the hypercharge direction, which always leaves nontrivial multiplicities for $X$, arising from off-diagonal components of the $SU(5)$ adjoint.

The expression (\ref{Pmulti}) provides a similar way to obtaining the spectrum 
for an associated spectral cover, where there is matter curve component not on $S$.
For example,
\begin{equation} \begin{split}
 n_{d^c}
 &= \textstyle \lambda \big(\sigma \cap (\eta-4c_1+x)+ (\eta-c_1 -x) \cap
 x \big) \cap  (4 \sigma - \eta +4 c_1 )|_S \\
 & =\lambda \Big(\big(-2\eta \cap (\eta-4c_1+2x)+ 4(\eta-c_1-x)\cap
 x\big) \cap \sigma \\
  & \qquad  +
 (\eta- c_1-x) \cap x \cap (\eta-4c_1) \Big)_S \\
  &= - \lambda \left(2 \eta \cdot (\eta-4c_1 +2x) + 4(\eta-c_1-x) \cdot x\right).
\end{split} \end{equation}
Here $|_S$ means just reading off the coefficient of $\sigma$. The matter curves for $d^c,l$ and $\nu^c$ differ by $t_5,t_6$, hence the same multiplicity
$$ n_{d^c}=n_l=n_{\nu^c}. $$
The other nontrivial spectrum is similarly obtained
\begin{equation} \begin{split}
 n_{D_2} = n_{h_u}  &= \lambda\left( 2\eta \cdot
 (\eta-4c_1-x)-4(\eta-x) \cdot (\eta-4c_1-x) \right), \\
 n_{D_1} = n_{h_d} &= -\lambda \left( 2\eta \cdot
 (\eta-2c_1-x)+4c_1 \cdot (\eta-x) \right), \\
 n_{\nu^c}   &= -\lambda \eta \cdot (\eta-4c_1+2x).
  \end{split}
\end{equation}
In the first line, we dropped the complex conjugate and instead we use the conjugate cycle having extra minus factor.
We could access the matter curves and multiplicities of the neutrinos beloinging to $\bf 16$ of $SO(10)$. However,
we cannot calculate the multiplicity of the $SO(10)$ singlet neutrinos $\nu_1,\nu_2$, since they are the fermionic partner of gauge moduli thus lives on the bulk of $C_q$.

The anomaly cancellation condition for the color $SU(3)$ reads
\begin{equation*}
  2 n_q - n_{d^c} - n_{u^c} + n_{D_2} - n_{D_1}= 0,
\end{equation*}
from which we have two choices
$$ x=0 \quad \text{ or } \quad 4x=5\eta-c_1. $$
This is from the four dimensional anomaly considerations. The former is required by the independent six dimensional conditions, so we will choose that, letting $d_0$ be a trivial section.

We have final spectrum
\begin{equation} \label{finmultipl}
 \begin{split}
 n_q & = n_{d^c} = n_{u^c} = n_l = n_{e^c}= n_{\nu^c} =
 -\lambda \eta \cdot(\eta-4c_1), \\
  n_{h_d} &= n_{h_u} = n_{D_1} = n_{D_2}
  =
 -2 \lambda \eta \cdot(\eta-4c_1).
 \end{split}
\end{equation}
We stress once more that the product is for the divisors in $S$,
meaning that this is native property of $S$. It is because, by
choosing $z$ as a section of the normal bundle $N_{S/B}$, we are
assuming the base of elliptic fibration $B$ is again $\P^1$ fiber over $S$.
For three generations of quarks and leptons (and Dirac neutrinos), we choose the base $S$ such that
\begin{equation} \label{3gencond}
 - \lambda \eta \cdot (\eta-4c_1) = 3.
\end{equation}
This condition is to be discussed in the following section. It follows
that we have six pairs of Higgs doublets, six pairs of colored
Higgses and twelve singlet neutrinos.
The number of $SU(2)$ doublets is even, so there is no $SU(2)$ global
anomaly.

In (\ref{finmultipl}), the same numbers of generations for different multiplets can be tracked by GUT relation. The spectral covers
$C_X$, $C_{q'}$ and $C_q$ are respectively related to $U(1)_Y$, $U(1)_M$ and $SU(4)_\bot$ `flavor' symmetry, among which we turned on the universal flux (\ref{su4flux}) only on $C_q$.
Because the commutant of $SU(4)_\bot$ in $E_8$ is $SO(10)$, we could expect
that the matter multiplicities obey the GUT relation. However
the matter fields and Higgs doublets are not related; they are minimally related by $E_6$.

Since there is no flux along $U(1)_Y$, or $F_Y = 0$, there is no anomaly and  Green--Schwarz mechanism does not make the corresponding gauge boson massive. Also $J \wedge F_Y = 0$ means that there is no Fayet--Illipoulos term.
Nevertheless, turning on $U(1)_Y$ flux would be interesting. Since then the flux does not obey the unification relation of $SU(5)$ GUT, 
we may take care of  doublet-triplet splitting problem.

Another nontrivial direction is to turn on non-universal flux \cite{DW3,Marsano:2009gv}. The traceless condition (\ref{tracelessproj}) also allows the flux of the type $(n_{1} - p^*_{m1} p_{m2 *})(C_{m2} \cap \sigma)$ or $(n_1 p^*_2 n_2 p^*_1)\rho$ for a two-cycle $\rho \in H^2(S,{\mathbb R})$.

\subsection{Base manifold}

The condition for three generations is Eq. (\ref{3gencond}). Recall that $c_1 = c_1(S)$ depend on the surface $S$, and $\eta = 6c_1 -t$ hence $-t=N_{S/B}$ depends on $S$ through the three base $B$, which we specified below (\ref{Cartans}). 
The good candidates are Hirzebruch surfaces $\mathbb{F}_n$ and their blowing-ups. We take del
Pezzo surfaces $dP_n$, defined by blowing $n$ points of $\P^2 =
\mathbb{F}_1.$ It is generated by one hyperplane divisor $H$ and $n(\ge 2)$
exceptional divisors $E_i$ satisfying the relation
\begin{equation}
 H \cdot H = 1, \quad H \cdot E_i =0, \quad E_i \cdot E_j = - \delta_{ij}.
\end{equation}
Consider again the integral condition. On $C_q$,
\begin{equation}
 c_1(C_q) = (2 (\sigma+c_1)-(\eta+4c_1))_{C_q} = 2 \sigma -\eta-2c_1,
\end{equation}
we obtain
\begin{equation}
 \begin{split}
 c_1(\N_q) &= -\frac12(2 \sigma+3c_1 + \eta) +
 \lambda(4\sigma-\eta+4c_1) \\
 & \textstyle = (-1+4\lambda) \sigma + \big(\frac12-\lambda \big) \eta + \big (\frac32 +  4\lambda \big )c_1.
 \end{split}
\end{equation}
It requires $4\lambda$ is an integer and the divisor class for the
other terms should belong to integer cohomology of $S$.
Taking
\begin{equation}
 \eta =4c_1 +H-E_1-E_2, \quad \lambda = 1
\end{equation}
we can achieve three generations \cite{DW1}.
The ampleness of
\begin{equation}
  t = 5H-E_1 -E_2 - 2 \sum_{i \ge 3}^n E_i,
\end{equation}
that $t \cdot x >0$ for any $x \in S$, is easily
checked, thus the cycle $S$ is contractable and the decoupling limit
exists \cite{BHV1}. One thing to note is, since the canonical class is
\begin{equation}
 c_1 = -K_S = 3H - \sum_{i=1}^n E_i,
\end{equation}
the even class condition for $\eta-c_1$ is always satisfied if $n$ is
even. Thus the minimal surface realizing the three generations is $S=dP_2$.
For the explicit construction of global geometry, we refer to Refs. \cite{Blumenhagen:2009yv,Collinucci:2009uh,Chen:2010ts}

The nonabelian gauge degree of freedom is described by M2 branes
wrapping 2-cycles $\{t_i\}$. We can expand four form field strength as
\begin{equation}
 G = \omega_i \wedge F_i
\end{equation}
where $\omega_i$ are (1,1)-harmonic basis, dual to $\{t_i\}$. 
Thus
 the
  $G$-flux has components $F_i$ along Cartan subalgebra of $E_8$. This
  is $\gamma$. And we have
\begin{equation}
 \frac12 \int_X G \wedge G = - \frac12 \int_S \pi_* \gamma^2.
\end{equation}
Tadpole cancellation is related to Euler number $\chi$ of
Calabi--Yau fourfold $X$
\begin{equation}
 \frac{\chi}{24}  = n_3 + \frac12 \int_X G \wedge G
\end{equation}
implying that we need $n_3$ three-branes to cancel anomaly.
Since nonabelian gauge symmetry enhancement takes place in the
singular limit, we have additional contributions to Euler number.
We can calculate it using generalized Pl\"ucker formula \cite{Andreas:1999ng,Andreas:2009uf}.

The other $E_8$ is located at the section at the infinity  $z \to
\infty$, which is totally disconnected to the subgroups $E_8$ at $S =
\{z = 0\}$. It can be partly broken, so that, for instance a small
nonabelian gauge group gives rise to gaugino condensation by
nonabelian group, contributing gravity mediation of supersymmetry
breaking. It will lower the number of three-branes.

\section{Outlook}

We conclude with an outlook about the low-energy theory.
We have obtained an F-theory derived model that is close to the Minimal
Supersymmetric Standard Model. Its gauge group $SU(3) \times SU(2) \times U(1)_Y$ is the commutant to the structure group $SU(5) \times U(1)_Y$ in $E_8$. The unification structure along $E_n$ series groups suggests that this group is unique. We can understand the corresponding singularity as a deformation of the $SU(5)$ singularity $\I_5^{\rm s}$. Upon breaking, the supporting surfaces for $SU(3)$ and $SU(2)$ are back-reacted from the original position of that for $SU(5)$, and the center-of-mass of brane stacks is responsible for the $U(1)_Y$, as in the perturbative description. 

We calculated the matter spectrum and its localizing curves using spectral covers. 
The intersections between spectral covers and branes for the gauge theory localize fields. They are identified by relating their positions with the field quantum numbers under the structure group. 
The conditions from six dimensional Green--Schwarz mechanism agrees with the consistent unification relation.

To distinguish electroweak Higgs bosons from lepton doublets, we need factorization of the spectral cover. A simple $\Z_4$ factorization is enough for it, and also forbids lepton and/or baryon number violating interactions up to an interacting scale. With a flux only on this `$\Z_4$'-cover, we can obtain three generations of quarks and leptons, with six pairs of electroweak Higgses and six pairs of colored Higgses. 
This flux is along the direction not violating $SO(10)$ GUT.
So, although we have constructed just the Standard Model group, the gauge group and the matter contents obey the unification relation of 
$SU(5)$ and $SO(10)$.

We have the standard doublet-triplet splitting problem, also tightly
related to the $\mu$-problem.
We can further elaborate the model, employing a different factorization and/or
fluxes. Nevertheless, already this model has desirable symmetries to shed light on a dynamical resolution.
Because of the $SU(5)_\bot\times U(1)_Y$ invariance, $\mu$-terms or colored  Higgs mass terms of the F-theory scale, close to the Planck scale, are forbidden. We expect some `standard solution' would break this symmetry and dynamically generate the mass matrix $m_D$ and $m_h$ at some intermediate scale.
Collecting the nonvanishing Yukawa couplings we have the superpotential close to that of MSSM,
\begin{equation*} \begin{split}
 W =&\ y_l l h_d e^c + y_u q h_u u^c + y_d q h_d d^c + m_{h} h_u h_d  \\
    &+ y_{\nu} l h_u \nu^c+ m_{\nu} \nu_M \nu_M^c +
     m_D D_1 D_2
  \end{split}
\end{equation*}
where $y$'s are matrix-valued Yukawa couplings.
The Higgs sector have flavor structure, so a phenomenological study in this sector, leaving only one pair of light Higgs doublets, would be interesting. Here the proton decay problem is milder than those in $SU(5)$ since we have no mixing of triplet Higgses with quarks.

The continuous version of matter parity $U(1)_M$ is well-known `$B-L$', the commutant to $SU(5)$ in $SO(10)$. The conventional expectation is it would be broken down to $\Z_2$ symmetry, becoming the {\em matter parity.}
Our minimal model does not have such a field, except the right-handed sneutrinos $\tilde \nu$ belonging to ${\bf 16}$ of $SO(10)$ doing a similar job. Their VEVs $\langle \tilde \nu \rangle$ break $U(1)_M$ down to a discrete symmetry $\Z_5$. There is an analysis along the context of heterotic compactifications with similar spectrum, where the renormalization group running generates desired potentials \cite{Ambroso:2009sc}.

The gauge couplings, contrary to GUT constructions, receive no contribution from $G$-flux since we do not have flux component along any visible Abelian gauge group, e.g. $U(1)_Y$. So the coupling unification is natural at the F-theory scale from the geometric embedding structure to $SU(5)$.

\subsection*{Acknowledgements}
The author is grateful to Deog Ki Hong, Jihn E. Kim, Joseph Marsano, Hiroshi Ohki, Taizan Watari and Piljin Yi for discussions and correspondences, and to KIAS, Pusan National University and Seoul National University for hospitality.
Especially he thanks to Tatsuo Kobayashi and Bumseok Kyae for the collaborations and reading the manuscript.

This work is partially supported by the Grant-in-Aid for Scientific
Research No. 20$\cdot$08326 and 20540266 from the Ministry of
Education, Culture, Sports, Science and Technology of Japan.

\end{document}